\documentclass[twocolumn]{article}
\usepackage{graphicx}
\usepackage{epstopdf}
\usepackage{ametsoc}
\usepackage{amsmath}

\catcode`\"=\active \let"=\"
\let\3=\ss

\def\C{{\bf C}}
\def\I{{\bf I}}	

\def\L{{\bf L}}

\def\Q{{\bf Q}}

\def\R{{\bf R}}

\def\B{{\bf B}}

\def\A{{\bf A}}
\def\F{{\bf F}}

\def\M{{\bf M}}

\begin{document}
\title{Drift wave-zonal flow dynamics}
\author{{\bf Brian F. Farrell} \\
Department of Earth and Planetary Sciences\\
Harvard University\\
Cambridge, MA 02138
\thanks{corresponding author: Brian Farrell, Harvard University, Department of Earth and
Planetary Sciences, Geological Museum 452, 24 Oxford Street, Cambridge, MA 02138.
email:farrell@seas.harvard.edu}\and
{\bf Petros J. Ioannou}\\
Department of Physics\\
National and Capodistrian University of Athens\\
Athens, Greece
}

%%\title{Emergence and equilibration of zonal jets in drift wave turbulence}
%\title{Drift wave-zonal flow dynamics}
%% Force line breaks with 
%\author{Brian F. Farrell} \email{farrell@seas.harvard.edu}
%\affiliation{%
%Department of Earth and Planetary Sciences, Harvard University\\
%24 Oxford Street, Cambridge, MA 02138, U.S.A.}

%\author{Petros J. Ioannou}
%\email{pjioannou@phys.uoa.gr}
%\affiliation{
%Department of Physics, University of Athens\\
%Panepistimiopolis, Zografos 15784, Greece
%}%

%\date{\today}
 \maketitle
 
\begin{abstract}
A remarkable phenomenon in turbulent flows is the spontaneous emergence of coherent large
spatial scale zonal jets.   Geophysical examples of this phenomenon include the Jovian
banded winds and the Earth's polar front jet.  In this work a comprehensive theory for the interaction of jets with turbulence,
Stochastic Structural Stability Theory,
is applied to the problem of  understanding the formation and maintenance of the zonal jets  that are crucial for enhancing plasma confinement in fusion devices. 
\end{abstract}

\section{Introduction}

Coherent jets that are not forced at the  jet scale
are often observed in turbulent flows with a familiar geophysical  example
being the zonal winds of the gaseous planets
\citep{Ingersoll-90}. This phenomenon of spontaneous jet formation
in turbulence has been extensively investigated in observational
and theoretical studies 
\citep{Balk-etal-1990, Panetta-93, Vallis-Maltrud-93, Cho-Polvani-1996, Huang-Robinson-98,  Farrell-Ioannou-2003-structural, Farrell-Ioannou-2007-structure, Diamond-2005, Connaughton-2009}
 as well as in laboratory experiments \citep{Krishnamurti-Howard-1981,Fujisawa:2008, Itoh:2007a, Itoh-2007b, Read-etal-2007, Mazzucato:1996, Holland:2006} .
The mechanism by which  these zonal flows form and  are maintained  is 
systematic organization of upgradient  eddy momentum flux in which the transfer of momentum occurs directly from the eddy field to the zonal flow without passing through intermediate scales, in contrast to the prediction of theories based on two dimensional turbulence cascades \citep{Nozawa-and-Yoden-97, Huang-Robinson-98, Ingersoll-etal-2004, Salyk-etal-2006, Kitamura-Ishioka-2007, Diamond-2005}.

Excitation of the eddies that give rise to zonal jets in turbulence can be traced either to predominantly external processes such as convection, as in the case of the Jovian
jets, or to predominantly internal  processes such as baroclinic growth,  as in the Earth's polar front jet.  However, maintenance of turbulence in a given flow is usually due to a combination of external and internal processes as for instance latent heat release associated with cumulus clouds injects external potential vorticity perturbations into the baroclinic turbulence of the  polar front jet.

Because of their self-regulating nature and interdependence drift wave turbulence and zonal flows behave as a single drift wave - zonal flow system (hereafter DW-ZF) \citep{Diamond-2005}.   In this system the drift wave perturbations  arise from the internal instability of the imposed density gradient, from sources external to the intrinsic dynamics of the drift waves and at a given scale  from transfer between scales by the internal quadratic nonlinear advection.  Because  these 
processes produce perturbations with short time and space scales compared to the 
time and space  scale of the jet, the associated eddy dynamics can be simulated using a Stochastic Turbulence Model  (STM) in 
which the nonlinear scattering and extrinsic excitation are modeled as stochastic
\citep{Farrell-Ioannou-1993d, Farrell-Ioannou-1996a, DelSole-Farrell-1996,  Newman-etal-97,  Zhang-Held-99, DelSole-04}.  
The STM provides an analytic method to
obtain the dynamics  of the quadratic statistics of a turbulent eddy field  associated with a given jet structure.  
Coupling a time dependent STM to an evolution equation for the jet  produces a dynamical system for the co-evolution of the jet and the self-consistent quadratic statistics of its associated turbulence; this is the method of Stochastic Structural Stability Theory  (hereafter SSST).   
The SSST system can be interpreted as  the dynamics of the ensemble mean jet and the ensemble mean associated turbulence in which the turbulence is modeled by the ensemble mean  perturbation linear dynamics 
with a stochastic approximation for the non-linear dynamics.  The solution for the eddy field is in terms of a covariance matrix from which can be obtained the  Gaussian probability density function
approximation for the  variance and  quadratic fluxes of the turbulence.
%The SSST system can be viewed as a computationally viable  approximation to the solution of the computationally unobtainable Fokker-Planck equation for a turbulent system.  
 The solution trajectory of the SSST equations often converge to a fixed point  state of balance between the
turbulence and the jet; however, limit cycles and chaotic solutions also occur
\citep{Farrell-Ioannou-2003-structural, Farrell-Ioannou-2007-structure, Farrell-Ioannou-2008-baroclinic, Farrell-Ioannou-2009-equatorial, Farrell-Ioannou-2009-closure}.  Chaotic trajectories of the SSST system correspond not to chaos of an individual turbulent state trajectory, which typically would be associated with a 
fixed point of the SSST system,  but rather to chaos of the ensemble mean turbulent state itself.  A familiar example of this type of chaos 
is the irregular  bursting behavior seen in drift wave turbulence \citep{Mazzucato:1996}.  Conceptually it is useful to view the SSST system as a computationally tractable approximation to a deterministically initialized Liouville system for the associated flow.

Interaction between the  zonal flow and its consistent field of turbulent eddies is nonlinear and can support multiple equilibrium states.  In many cases these equilibrium states  arise from and can be traced by continuation in a parameter to a bifurcation of the   coupled DW-ZF system.  In the case of  the equilibrium state with no mean density gradient and no initial zonal flow the zonal flow  forming bifurcation arises  as a function of a parameter controlling turbulence intensity as  an emergent  instability  of the SSST system intrinsic to the interaction between
the zonal flow  and the turbulence.  One may think of a  perturbation zonal flow organizing the surrounding turbulence to produce a momentum flux divergence that amplifies that perturbation zonal flow.  The particular perturbation  zonal flow structure that organizes the turbulence to exactly amplify  its own structure is obtained as an eigenfunction of the perturbation SSST system linearized about a marginally stable SSST equilibrium.  
This instability equilibrates at finite amplitude and this finite amplitude SSST equilibrium, consisting of the zonal flow  and associated consistent  eddy field,  can be connected by continuation in an appropriate parameter, such as the density gradient, to nearby finite amplitude equilibrium states.  

In addition to simply continued equilibria there also exist equilibria that are isolated  to variation of a  given parameter as for instance a strong zonal flow  equilibrium exists at a moderate density gradient and turbulence intensity  that can not be connected by continuation starting from a weak zonal flow equilibrium at a small density gradient.  However,  external turbulence excitation can be used as a control parameter   to promote  the system to such an isolated equilibrium state.  In addition to parameter control we may also perturb the zonal flow to promote the system to an   isolated equilibrium state.  Promoting the DW-ZF system to different regime states by parameter control is analogous to instigating a laminar/turbulent transition in shear flow turbulence where the Reynolds number is the control parameter.

An equilibrium state of balance between a zonal flow and its associated field of  drift wave turbulence   requires that the momentum flux divergence arising from the turbulence precisely balance the zonal flow  momentum loss to friction, if any.  The requirement of a precise balance between zonal flow forcing and dissipation, if any,  is far more demanding than that the shear associated with the jet simply suppress the turbulence while the turbulence  during  the suppression process produces up-gradient momentum flux.  The remarkable fact is that the turbulence, which depends on the zonal flow, and the zonal flow, which depends on the turbulence, mutually adjust to produce balanced states.  Having the SSST analytic dynamics of the DW-ZF system   allows us to predict  parameter values for which  robust equilibrium DW-ZF regimes  are maintained,  to predict parameters values for which  time dependent periodic and chaotic DW-ZF regimes occur, to predict transition between these regimes when two regimes exist at the same parameter values,  and ultimately to predict the breakdown of the zonal flow regime.

Closer inspection of the density transport  mechanism reveals that  the observed and simulated DW-ZF  equilibrium  jet density transport suppression can not be understood using the concept of effective diffusion  \citep{Sanchez:2009}.  In effective diffusion theory it is assumed that  transport of a passive scalar is  proportional to the  scalar gradient with coefficient $D_{eff} = {v}{l}$ in which $v$ and $l$ are the characteristic velocity and spatial correlation  scales of the turbulence.  Transport can vary either due to changes in the characteristic velocity   or in  the eddy correlation scale.   In this work we solve for the correlation between velocity and density fluctuations directly  
revealing turbulent transport both up and down the mean gradient,  in agreement with observations and simulations \citep{Shats:2000, Holland:2006}, and implying the density transport process in drift wave turbulence is not diffusive in nature. Instead we find that large scale coherent structures rather than small scale 
eddy diffusion are responsible for density transport \citep{Bos-etal-2008}. 

Closer inspection of the dynamics of the interaction between perturbations and zonal flows  reveals that understanding reduction  of turbulence variance by zonal flows through the concept of shear suppression by zonal flow advection is incomplete.  Shear suppression  has roots in WKB theory and the concept of a continuous spectrum of advected harmonic waves.  However, to properly understand perturbation dynamics in jets a full wave solution must be obtained because the perturbation dynamics supports a complete set of large scale coherent  modes that are in general not orthogonal and among which exists a subset that is potentially unstable.  Interaction between the zonal jet and the eddy field systematically  stabilizes  these modes \citep{Ioannou-Lindzen-1986, James-1987,
Roe-Lindzen-1996}  during the establishment of  a statistical equilibrium.  Moreover, the non-normal equilibrium  jet dynamics supports a subset of stable structures that produce robust growth under internally and externally imposed excitation.  These Stochastic Optimal (SO) perturbations \citep{Farrell-Ioannou-1996a}  comprise a small subset of structures but these are the structures responsible for  growth of perturbations due to interaction with the zonal shear and density gradient.
Using SSST we solve for the complete normal mode eigenstructure of the equilibrium  jet as well as  the  SO and EOF (Karhunen-Loeve) decomposition of the ensemble mean turbulence variance and cross variance in  the velocity and density fields.  This analysis provides full information on the structure and dynamics of the perturbations responsible for producing the turbulence variance and fluxes.

The mechanism of  jet formation in plasmas can be studied for turbulence arising from external, internal, or a combination  of sources.  The Charney-Hasegawa-Mima (C-H-M) equation provides the simplest model system as it uses only external turbulence excitation.   Zonal jet formation in this model is identical to that in the  equivalent barotropic vorticity equation \citep{Farrell-Ioannou-2007-structure}.  However, because in the DW-ZF problem there exists an internal instability associated with the density gradient  this problem is more comprehensively modeled using the modified Hasagawa-Wakatani (H-W) equations which describe plasma dynamics in a 2D slab model.  These equations are similar, although not identical, to the  baroclinic two-layer model \citep{Farrell-Ioannou-2008-baroclinic}.  In this work we use the H-W equations  to study DW-ZF dynamics.

The SSST equations incorporate a stochastic turbulence model but these equations are themselves deterministic and autonomous with dependent variables the zonal flow and the ensemble mean covariance of the turbulence.  It follows that the perspective on stability provided by these equations differs from the more familiar perspective based on the perturbation stability of the zonal flow.  In fact the primary bifurcation in these equations has no counterpart in  zonal flow stability analysis; it is rather a cooperative instability  in which the perturbation zonal flow  organizes the background turbulence to produce flux divergences configured to amplify the jet leading to an emergent turbulence-zonal flow instability that need not coincide with perturbation instability of the jet.  The SSST equations are the nonlinear ensemble mean dynamics for the DW-ZF flow system and this system in many cases supports equilibration of the emergent jets and their consistent turbulence fields at finite amplitude.  These finite amplitude equilibria in turn lose structural stability as a function of parameters and this instability is associated either with bifurcation to another equilibrium or to loss of a stable equilibrium state.   While it is true that loss of modal jet stability by an equilibrium state as a function of a parameter also implies loss of structural stability, the converse is not true.  For this reason bounds on zonal jet amplitude based on  modal instability of the jet are not tight and can often be improved by analysis of the structural stability of the jet.

A gradient driven flow with constant density gradient is assumed in the examples below for simplicity although the particle flux is calculated and could be used with an appropriate density gradient forcing parameterization to obtain equilibria in which the density gradient participates in the equilibration.  However, as equilibrium is approached the fluxes are typically suppressed implying long time scales for changes in the equilibrium density gradient by flux divergence and the likelihood that external driving mechanisms dominate density gradient variation.

\section{Formulation}

\subsection{The Hasegawa-Wakatani drift wave turbulence equations}
%Consider the H-W equations which represent the ion density and vorticity in a slab geometry 
%identifying the x coordinate with the radial direction and the
%y coordinate with the poloidal direction. 
We use the modified Hasegawa-Wakatani (H-W)  equations \citep{Numata-etal-2007}.  These equations model the turbulence of  the edge region  of a tokamak plasma with fractional density decreasing in the radial direction, $x$, at
a constant rate $\kappa$, so that $n(x)=n_0 e^{-\kappa x}$, and  in a constant background magnetic field $\B=B_o \bf {\hat z}$ in the toroidal, $z$,  direction.   The H-W equations govern the dynamics  of the 
electrostatic potential $e \phi / T_e$ and the ion density $n/n_0$ in a cartesian approximation of the radial-poloidal,  $x-y$, plane. 

The ion vorticity,  $\zeta=\Delta \phi$, and the density fluctuations, $n$, solve following  \cite{Numata-etal-2007}:
\begin{subequations}
\begin{align}
&&\partial_t \zeta + J[\phi,\zeta]  = \alpha(\phi'-n') + \nu \Delta \zeta ~,
\label{eq:HWa} \\
 &&\partial_t n + J[\phi,n] =   \alpha(\phi'-n') -\kappa \partial_y\phi+\nu \Delta \zeta
\label{eq:HWb}
\end{align}
\end{subequations}
with Jacobian $J[f,g] \equiv (\partial_x f)( \partial_y g)-(\partial_y f) (\partial_x g)$. The fields are decomposed into zonal means  and departures from  zonal means:
\begin{equation}
\phi = \overline{\phi} + \phi'  ~~~, ~~n = \overline{n} + n'~,
\end{equation}
with the zonal mean, denoted by a bar, defined as the mean in the poloidal direction $y$:
\begin{equation}
\overline{f} = L_y^{-1} {\int_0^{L_y} f(x,y,t) dy} ~.
\end{equation} 
The  flow velocities are:
\begin{equation}
u= - \partial_y \phi ~~,~~v=\partial_x \phi~.
\end{equation}

%\vskip .2in

The parameter $\alpha$  controls
the strength of the  electron resistivity that couples the electrostatic field with the ion density
perturbations.  For $\alpha=0$  equation (\ref{eq:HWa}) for the electrostatic potential $\phi$ corresponds to the hydrodynamic 2D vorticity equation 
while the density equation (\ref{eq:HWb}) corresponds to the advection-diffusion
of  $n'$ as a passive scalar in the presence of a mean fractional radial density gradient $-\kappa$. 
In the limit  $\alpha \rightarrow \infty$ the density and electrostatic field couple rigidly and obey the Charney-Hasegawa-Mima  equation   \citep{Hasegawa-1978}.  The dynamics of this  equation, which  governs the formation of zonal flows in both the GFD and the plasma context,  has been studied in recent theoretical work on zonal flow generation  \citep{Balk-etal-1990, Connaughton-2009, Farrell-Ioannou-2003-structural, Farrell-Ioannou-2007-structure}. 
Hereafter, we treat the more general quasi-adiabatic case with $\alpha=1$, and allow for instability by including an ion density gradient,  $\kappa$, which will be treated as a variable parameter.

In the nondimensionalization of the equations lengths are scaled by the Larmor radius 
$\rho_s = \sqrt{T_e/m_i}  \omega_{ci}^{-1}$ and time by  the  electron cyclotron frequency    $\omega_{ci}=  e B_0 /  m_i$. 
A typical Larmor radius, $\rho_s = 1 ~mm$, is obtained for a magnetic field $1~ T$ and  electron temperature  $T_e=95.6~eV$; also for these values
 $\omega_{ci}^{-1}=  10^{-8} \rm~ s / rad$ and  the
corresponding velocity scale $\rho_s \omega_{ci}$ is $95.6~ km /s$.  The channel is taken doubly periodic in both $x$ and $y$.

The zonal average of (\ref{eq:HWa}) gives the equation for the zonal jet:
\begin{equation}
\partial_t \overline{v} =  - \overline{u' \zeta'  } -r_m \overline{v}~.
\label{eq:meana}
\end{equation}

Where $\overline{v}=D \overline{\phi}$ and $D\equiv \partial_x$. The zonal flow is damped linearly at the
mean collisional damping  rate, $r_m$, which will typically be taken to be $r_m=10^{-4} $ although we will also present results in the 
collisionless limit, $r_m=0$.

The nonzonal components obey the equations:
%\begin{widetext}
\begin{subequations}
\begin{align}
 && \partial_t \zeta'  = -\overline{v} \partial_y \zeta'  +( D^2\overline{v} ) \partial_y \phi' + \alpha (\phi' - n') 
+ \nu \Delta \zeta'  + F(\zeta') \label{eq:perta}\\
&&  \partial_t n'= -\overline{v} \partial_y n'  -\kappa \partial_y \phi'+ \alpha (\phi' - n') 
 +\nu \Delta n'  +F(n')~. \label{eq:pertb}
 \end{align}
\end{subequations}
%\end{widetext}
with nonlinear scattering term:
 \begin{equation}
F(f)= - \partial_x( u' f' - \overline{u' f'} ) - \partial_y ( v' f' - \overline{ v' f'} )~.
\label{eq:scatter}
\end{equation}
These equations can sustain turbulence without external forcing  due to the radial density flux,
 $\overline{u' n'}$, in the presence of the mean density gradient \citep{Numata-etal-2007}.
We now briefly review
the energetics of these equations. The total energy, $E$, 
is the sum of the zonal mean kinetic energy:
\begin{equation}
\overline{E} = \frac{1}{2} \int_0^{L_x} {\overline v}^2 dx ~,
\end{equation}
and the eddy energy:
\begin{equation}
 E' = \frac{1}{2} \int_0^{L_x}\left (  |\nabla \phi |^2  + n^2 \right ) dx ~.
 \end{equation}
From the zonal mean  equation (\ref{eq:meana}) we obtain:
\begin{equation}
\frac{d \overline{E}}{d t} =  \Gamma_e - 2 r_m \overline{E}~,
\end{equation}
where
\begin{equation}
\Gamma_e=- \int_0^{L_x} \overline{v}~ \overline{u' \zeta'} dx
\end{equation}
is the  time  rate of change of the zonal mean  energy due to the eddy induced mean zonal  acceleration $- \overline{u' \zeta'}$. 
Similarly, we obtain from the perturbation equations (\ref{eq:perta},\ref{eq:pertb}):
\begin{equation}
\frac{d {E'}}{d t} =  - \Gamma_e +   \Gamma_n - \Gamma_{\alpha} - \Gamma_{\nu} + F~,
\label{eq:pertE}
\end{equation}
where  
\begin{equation}
\Gamma_n = \kappa \int_0^{L_x} \overline{u' n'} dx~,
\label{eq:Gn}
\end{equation}
is the rate of perturbation energy gain due to perturbation density flux down the mean density gradient. This term 
provides  the internal energy source for the  turbulence.
The term
\begin{equation}
\Gamma_{\alpha} = \alpha \int_0^{L_y} (\phi - n)^2 dx ~,
\label{eq:Galpha}
\end{equation}
corresponding to resistive coupling is always dissipative as is the diffusion:
\begin{equation}
\Gamma_{\nu} =\int_0^{L_y} \left ( |\nabla \zeta'|^2 +  |\nabla n'|^2 \right ) dx  ~.
\end{equation}
The term, $F$, is the rate of  energy input by external excitation.  This   external energy input rate is constant if the excitation  is delta correlated and state independent.

\subsection{The SSST system governing DW-ZF dynamics}

We parameterize   the nonlinear scattering term (\ref{eq:scatter}) in the eddy equations (\ref{eq:perta}, \ref{eq:pertb}) by stochastic forcing, which is the STM closure \citep{Farrell-Ioannou-1993e, Farrell-Ioannou-1994a}. 
The STM  accurately simulates both the  structure of the eddy field  and of the quadratic fluxes in shear turbulence including that of the Earth's atmosphere, which is a particularly well observed  turbulent medium \citep{Farrell-Ioannou-1995,DelSole-96, Whitaker-Sardeshmukh-98, Zhang-Held-99, DelSole-04}.  

We  represent the  perturbation fields using Fourier components in  the poloidal direction, $y$: 
\begin{equation}
\phi'= \sum_k \hat \phi_{k}(x,t) e^{i k y}~,~n'= \sum_k \hat n_{k}(x,t) e^{i k y}~,
\end{equation}
and discretize the equations  in the radial direction, $x$, so that the state $\psi_k= [ \hat \phi_k , \hat n_k]^T$
is prescribed by the values, for each Fourier component, of the electrostatic potential and the perturbation density 
on an equally spaced grid.  
Under the simplifying assumption that the stochastic forcing has sufficiently short temporal correlation that it can be approximated as white noise, the  second moment statistics of the fluctuating field, $\psi_k$,  are fully described by the covariance matrix 
$\C_k= < \psi_k \psi_k^{\dagger} >$ ( $<\cdot >$ denotes ensemble averaging) which evolves according to the deterministic Lyapunov equation:
 \begin{equation}
  \frac{d \C_k }{d t}~ =~\A_k(\overline v) \C_k~+\C_k \A_k^{\dagger}( \overline v)~+~\Q_k~,
   \label{eq:Lyap1}
  \end{equation}  
  in which $\Q_k$ is the covariance representing the ensemble average distribution of the stochastic forcing in the radial direction 
   \citep{Farrell-Ioannou-1996a} and   $\A_k(\overline{v})$ 
  is the linear  operator  in
  (\ref{eq:perta}, \ref{eq:pertb}) which depends affinely on the  zonal flow $\overline{v}(x,t)$.  
 If  $\Q_k$  represents scattering by the advective nonlinearity rather than external sources of excitation then a dissipation 
can be added to the linear operators to ensure that no net energy is introduced into the system (because the nonlinear terms only redistribute energy).  Also,  $\Q_k$   can be made an appropriate function of the amplitude of the
 perturbation variance in order to accurately parameterize the quadratic nonlinearity of the advective Jacobian.
More comprehensive closures of this sort have been used in other contexts
 \citep{DelSole-01b,Farrell-Ioannou-2009-closure}; however,  it is sufficient for our present purposes to use the 
simplest parameterization  in which the system  is stochastically excited with state independent forcing and the behavior of the system is investigated  as a function of the amplitude of this excitation.

The Lyapunov equation (\ref{eq:Lyap1}) determines  $\C_k$ and this covariance  in turn determines the ensemble mean vorticity flux:
\begin{eqnarray}
\left < {u' \Delta \phi'} \right > &= &\sum_k \frac{1}{2} \Re  \left [ - i k \left < \hat \phi_k  \Delta_k \hat  \phi_k^{'*} \right > \right ]   \nonumber\\
& =& \sum_k \frac{k}{2} \Im \left [ {  \rm diag} ( \C_k \Delta_k^{\dagger} ) \right ] ~.
\end{eqnarray}

However, it is the zonal  mean vorticity flux that  appears in  the zonal flow equation (\ref{eq:meana}) but 
under the ergodic assumption  the zonal mean can be replaced by the ensemble mean: 
\begin{equation}
\left < {u' \Delta \phi'} \right > = \overline{u' \Delta \phi'} ~.
\end{equation}
This  requires that there be many independent realizations of eddy activity  in the poloidal direction 
and in that limit we obtain  the ensemble mean  equations:
\begin{subequations}
\begin{align}
\partial_t \overline{v} =   - \sum_k \frac{k}{2} \Im \left [ {  \rm diag} ( \C_k \Delta_k^{\dagger} ) \right ]  -r_m \overline{v}~\label{eq:S3Ta}\\
\frac{d \C_k }{d t}~ =~\A_k(\overline v) \C_k~+\C_k \A_k^{\dagger}( \overline v)~+~\Q_k~.
\label{eq:S3Tb}
\end{align}
\end{subequations} 
 The equation for the turbulence  covariance,
(\ref{eq:S3Tb}),  and the  equation for the  mean
zonal flow, (\ref{eq:S3Ta}),  together comprise a closed system for the evolution of the zonal flow under the
influence of its consistent field of turbulent eddies. 
Although the effects of the ensemble mean turbulent fluxes are
retained in this system, the fluctuations associated with the
turbulent  eddy dynamics  are suppressed  and the
dynamics becomes autonomous and  deterministic. These SSST equations can be interpreted as the dynamical equations for  the evolution of the quadratic (Gaussian)  approximation to the ensemble mean probability distribution of the   turbulent  DW-ZF  system.  
%From this perspective SSST can be viewed as a computationally accessible approximation to the Fokker-Planck equation for this system.  
This concept invites novel perspectives such as that of  chaos of the ensemble mean state of a turbulent system as distinct from chaos of a realization of the system.  We show examples of ensemble mean chaos in DW-ZF turbulence below.
However, the SSST system trajectory is often not chaotic but instead asymptotes to a fixed point equilibrium and in these
cases the dynamics of DW-ZF equilibria  emerge with great clarity in the SSST system. 
As another  illustration of the insight provided by this  system we note that zonal jets arise in SSST as  easily analyzed linear instabilities.
This jet forming instability  is an example of a new class of instability in fluid dynamics; 
it is an emergent instability that arises essentially from the interaction between the zonal flow and the turbulence.

\subsection{Parameters}

Unless otherwise indicated  calculations were performed  with $64$ points in the $x$ direction  and 8 harmonics in the 
$(y)$ direction comprising
wavenumbers  $k= [ k_0,3k_0,5k_0,7k_0,9k_0,11k_0,13k_0,15k_0]$ with $k_0=0.15$ 
in a doubly periodic channel  with  $L_y =2 \pi/ k_0$ and  $L_x=L_y/4$. 
The stochastic forcing is taken to have an identity covariance in vorticity corresponding to a one grid point correlation, and is normalized 
so that the energy input by the stochastic forcing is the same for all zonal wavenumbers.  The excitation  of the electrostatic field and the density field
is correlated in order to facilitate  the  adjustment of the two fields (similar results are  obtained using uncorrelated forcing). The amplitude of the stochastic forcing
is given in terms of the equivalent $u_{rms}$  velocity that would be maintained by the forcing   with no zonal flow  and with $\kappa=0$.
Dissipation parameters used are  $\nu = 10^{-2}$, $\alpha = 1$ and $0 \le r_m \le10^{-4}$.

\section{Results}
\subsection{Formation of zonal jets starting from a non-equilibrium state}

The starting point for a systematic investigation of DW-ZF dynamics is the nonlinear SSST system initiated in a state lying on its  attractor. However, the system is commonly thought of as being initiated far from its attractor  in a state of high turbulence intensity but without the corresponding finite amplitude equilibrium jet.
There then ensues a rapid adjustment process in which the system builds a jet corresponding to the turbulence and in the process places the system on the SSST attractor. 
In order to study this adjustment process consider the example of the turbulence field associated with a single poloidal wavenumber in equilibrium with a strong stochastic excitation but without its consistent  zonal jet.  The turbulence field is that obtained from the stochastically excited STM but without coupling to the 
zonal flow equation.  If the zonal  flow equation is coupled to the STM at this point to form the interactive SSST system there ensues  rapid formation of a consistent zonal jet.  We show this rapid development of a jet from a strong  initial turbulent field at the single
global wavenumber $m=5$ in Fig. \ref{fig:1} for $\kappa=0$ and no stochastic excitation. We also show the unstable case with $\kappa=1$ and  
in both cases  the development of
the zonal jet is rapid because of the feedback between the eddies and the growing zonal jet. 
If as an experiment the eddies are required to develop on a fixed jet structure  that is not continuously modified by their dynamics then the resulting 
fluxes build 
the jet much more slowly revealing that  rapid jet formation is due to the cooperative DW-ZF interaction.   The build up of the jet and the subsequent suppression of the eddy energy  
occurs  due to  
shearing of the eddy field by the jet, a process discussed  by \cite{Diamond-2005} and that is 
seen  both in simulations \citep{Numata-etal-2007} and observations.   Because the vorticity flux is proportional to the shear 
\citep{Farrell-Ioannou-1993-unbd, Farrell-Ioannou-2009-equatorial} the rate of increase of the shear is proportional to the shear 
and the eddy variance, so that if the eddy variance is constant exponential growth of the shear results but if 
the eddy variance is also growing during this phase an exponential growth ensues with time increasing exponent.  

Similar development occurs when there is stochastic forcing and as a result the turbulence has a full spectrum. 
An example with $\kappa=1 $ of jet  emergence from small amplitude random initial conditions in an unstable flow  with 
substantial stochastic excitation is shown  in Fig. \ref{fig:2} and the process of  its approach to equilibrium is shown in Fig. \ref{fig:3}. 
The eddy induced zonal acceleration  reaches its peak 
during this initial development (cf. Fig. \ref{fig:2}d). The  rapid suppression of the  eddy variance
(cf. Fig. \ref{fig:2}c) is caused 
by energy transfer to the zonal  flow and by  increased dissipation,
$\Gamma_{\alpha}$, due  to increased disequilibrium of the electrostatic field $\phi'$ and 
the perturbation
ion density fluctuations $n'$   [cf.  the energetics equation
(\ref{eq:pertE})].  

After the initial development of the zonal jet by the mechanism of anti-diffusive shear momentum transport as described above there follows a  period of adjustment in which the SSST system  attempts to stabilize the
zonal flow and to establish, if the parameters allow it,   a steady state equilibrium  corresponding to a fixed point of the SSST equations. 
This stabilization process 
is shown in Fig. \ref{fig:3}  as it sequentially stabilizes the perturbation operator $\A_k$.   As the jet adjusts to equilibrium during this phase the flow is dominated by large structures and the adjustment has a full wave modal character unlike during  the initial period of jet formation from a state far removed from the system attractor in which the dynamics is shear wave anti-diffusion dominated being associated essentially with rapid distortion of the initial perturbation field.

\subsection{Structural instability of the zero zonal flow state as a function of the amplitude of the stochastic excitation   in the absence of drift wave instability, $\bf \kappa=0$.}

We turn now to dynamics on the attractor of the SSST DW-ZF system
and first study the case $\kappa=0$ in which there is no drift wave instability and eddy variance is maintained solely by external excitation.
In the absence of zonal flow the SSST equations (\ref{eq:S3Ta}, \ref{eq:S3Tb}) are translationally invariant  in the radial direction  and  the vorticity flux $\overline{u' \zeta'}$  vanishes 
and as a result the  zero zonal flow is an equilibrium of the SSST equations  for any stochastic excitation and associated turbulence level. 
The SSST equations can be linearized about this zero state $\overline{v}_E = 0$ and the eddy covariance that corresponds to a chosen  stochastic excitation of this zero state, 
$\C_{kE}$, obtained from the steady state Lyapunov equation.  About this state perturbation equations can be obtained for the perturbation zonal velocity, $\delta \overline{v}$, and perturbation eddy covariances, $\delta \C_k$, 
in the form:
\begin{equation}
\left [
\begin{array}{cc}
  \delta \overline{v}   \\
\delta \C_k
\end{array}
\right ]
=
\L(\overline{v}_E, \C_{kE})  
\left [
\begin{array}{cc}
  \delta \overline{v}   \\
\delta \C_k
\end{array}
\right ].
\end{equation}

The growth rate and structure of the most rapidly growing eigenmode of $\L$ provides insight into the mechanism of  zonal jet emergence and equilibration in turbulence\citep{Farrell-Ioannou-2003-structural, Farrell-Ioannou-2007-structure}. Zonal jets arise  as finite amplitude nonlinear equilibria proceeding from the most rapidly growing eigenmode of  
$\L$ linearized about the zero state.  Note that this jet forming instability does not in general coincide with loss of stability of the $\A_k$ operators which determine the stability of a finite amplitude   zonal flow to eddy perturbation.  

The SSST system can be linearized about finite amplitude SSST equilibria as well as about the zero state and the bifurcation structure about these finite equilibria can be examined as a function of parameters to determine e.g. the circumstances under which jet breakdown occurs.  It should be noted in this context that equilibria of the SSST system are necessarily perturbation stable.  Consider as an example the SSST stability of the  zero zonal flow state,  $[\C_{kE}, \overline{v}_E=0]$, with $\kappa=0$.  In this case  the  zonal jet emerges as increase in stochastic excitation, $\Q_k$, causes the turbulence level to exceed a threshold at which point  $\L$  becomes SSST unstable.
As the amplitude of the excitation, $\Q_k$, is increased further this bifurcation connects to finite amplitude equilibria  
in which the eddies maintain finite amplitude zonal jets.  The  bifurcation diagram of this zonal flow as a function of excitation amplitude  is shown in Fig. \ref{fig:4}a,c together with the associated non-linearly equilibrated zonal jets.  For weakly supercritical excitation the structure of the zonal  flow is nearly that of the most unstable mode of the $\L$ operator but as  the excitation increases 
the velocity of the zonal flow asymptotes to a constant structure  as shown in  Fig. \ref{fig:4}d. 

We can understand an important aspect of the dynamics of this  asymptotic structure by noting that as the stochastic excitation increases  the zonal flow acceleration associated with the ensemble mean Reynolds stress divergence:  
\begin{equation}
< \overline{  u' \zeta' }  > = \frac{1}{2} \sum_k \Re \left ( \hat{u}_k \hat {\zeta}_k \right )~,
\end{equation}
is comprised of a sum of low  zonal wavenumber fluxes that decelerate the jet   and high wavenumber fluxes that accelerate the jet.  As excitation and turbulence level increase  the vorticity flux of each component of this sum increases while the sum tends to the small residual required to balance the zonal flow dissipation because  the  low wavenumber  downgradient and high wavenumber  upgradient contributions  very nearly cancel \citep{Farrell-Ioannou-2009-equatorial}. This dynamic can be seen  in Fig. {\ref{fig:5} in which the structure of the vorticity fluxes  associated with the equilibrium
jet in Fig. \ref{fig:4}d is shown. In Fig. \ref{fig:5}a it is seen that  wavenumbers $m=1,3,5$ oppose the jet   and nearly cancel the upgradient contribution from the higher wavenumbers. This cancellation becomes all the more  complete as the excitation increases and the equilibrium  zonal flow assumes asymptotic form. 
Because the total vorticity flux vanishes in the collisionless limit, $r_m=0$,  these equilibria  are  also the equilibria in this limit (as shown in Fig. \ref{fig:5}b). 
This demonstrates that in turbulence with vanishing collisional damping of the zonal flow
there are non-vanishing  equilibria  that  are independent of the turbulence intensity  and have the  universal structure  shown in Fig. \ref{fig:4}d.  It should be noted that while  this asymptotic zonal flow
does not depend on the turbulence intensity for a fixed spectrum of excitation it is sensitive to the spectral distribution because the fluxes are upgradient for high wavenumbers and downgradient for low wavenumbers.  
For example, if only low wavenumbers are excited no finite equilibria result 
for any $r_m$ as all fluxes  oppose the jet.  Conversely, if only the high  wavenumbers are excited equilibria arise for $r_m \ne 0$ associated with upgradient fluxes but in this case the equilibrium  zonal flow  increases secularly with excitation increase until the jet became structurally  unstable. 

We note, in the examples shown and in agreement with observations and simulations, that the kinetic energy is concentrated in the energy of the zonal jet while the eddy kinetic energy is greatly suppressed (cf. Fig. \ref{fig:6}). 

\subsection{Zonal flow equilibria as a function of the amplitude of stochastic excitation in the presence of drift wave instability, $\bf \kappa>0$.}

Introduction of unstable density stratification $\kappa >0$ makes the zero state perturbation unstable and necessarily structurally unstable for any stochastic excitation. These unstable eddy perturbations augment the turbulence and facilitate formation of zonal  jets. 
An example with   $\kappa=1$  of jet emergence from small amplitude random initial conditions in an unstable flow  with substantial stochastic excitation are shown  in Fig. \ref{fig:2} and Fig. \ref{fig:3}. 
%The development of the zonal jet is rapid because of the feedback between the eddies and the growing zonal jet. 
%If as an experiment the eddies are required to develop on a fixed jet structure  that is not continuously influenced by their dynamics then the resulting fluxes build 
%the jet much more slowly.  The build up of the jet and the suppression of the eddy energy
%happens on the jet advective time scale and occurs by the perturbations rendering their momentum
%to the mean flow by 
%shearing over the existing shear flow, a process discussed extensively in this context by \cite{Diamond-2005} and is 
%observed both in simulations \citep{Numata-etal-2007} and observations.  The eddy induced mean acceleration also  reaches its peak 
%during this initial development (cf. Fig. \ref{fig:4}d). The  rapid suppression of the  eddy variance
%(cf Fig. \ref{fig:4}c) is caused 
%by transferring eddy energy to build the mean flow and by the  increased dissipation  implied by
%$\Gamma_{\alpha}$  from increased disequilibrium of the electrostatic field $\phi'$ and 
%the perturbation
%ion density fluctuations $n'$   $\Gamma_{\alpha}$ (cf  the energetics equation
%(\ref{eq:pertE})).

The radial distribution at various poloidal wavenumbers  of the equilibrium particle flux and of the 
acceleration by the Reynolds stress  are shown in Fig. \ref{fig:7}a,c.  The eddy induced acceleration of the zonal flow   by  small wavenumber eddies is downgradient,
opposing  the jet,  while the acceleration due to larger wavenumbers is upgradient, as for the case $\kappa=0$. This cancellation implies,
as for the case with $\kappa=0$, that the equilibrium flow asymptotes to a fixed structure as the amplitude of the forcing increases and that this asymptotic flow
is also the equilibrium flow in the collisionless limit, $r_m =0$.  Similar equilibria with  zero collisional damping have also been seen in turbulence simulations \citep{Lin-etal-2000}.
The  asymptotic equilibrium flow shown in
Fig. \ref{fig:8} is found to depend only weakly on $\kappa$. For $r_m=0$ this is the universal equilibrium flow 
for all forcing amplitudes and for all $\kappa$. However this equilibrium is structurally unstable for large values of $\kappa$, as will be discussed.

The eddy kinetic energy peaks at the gravest poloidal scale, $m=1$. It is important to note that it is at large scales  that most of the
eddy energy resides and also  it is the  large scales  that are responsible for the particle flux (the particle flux peaks for $m=3$ as shown in Fig. \ref{fig:7}b) as
is  also found in turbulent simulations \citep{Bos-etal-2008}.
The dominance of  large scales in the eddy variance and fluxes is consistent with these scales  being the least damped 
(cf Fig. \ref{fig:3}), however the eddy structure does not assume the structure of the least damped modes. We show 
the structure of the least damped mode and the distinct structure of the top EOF of the covariance matrix (this is the eigenfunction associated to the largest eigenvalue of $\C_k$) for the gravest poloidal scale $m=1$ in  Fig.  \ref{fig:9}e,f. We also show the  first stochastic optimal which is the
structure of the excitation that would produce, if the flow were forced stochastically with this structure at unit amplitude,   the highest  eddy energy at statistical 
equilibrium \citep{Farrell-Ioannou-1996a}. The difference in the structure of the top EOF, the least damped mode and the stochastic optimal reveal the degree of non-orthogonality of the modes of the operator which is related to their non-normality as these would be identical if the system were normal \citep{Farrell-Ioannou-1994b, Ioannou-1995}. 

The non-normality of the H-W system is central to  its dynamics.  In order to appreciate its role consider  the frequency spectrum of the total eddy variance resulting from excitation unbiased in time and space  of the linearized H-W equations. This can be  obtained by Fourier transforming the perturbation equations 
 (\ref{eq:perta}, \ref{eq:pertb}) written in the form:
\begin{equation}
\frac{d \hat \psi_k}{ dt} = \A_k  \hat \psi_k + \F \eta
\end{equation}
where $\eta$ is Gaussian white noise and $\F$ gives the radial structure of the forcing related to the  noise covariance $\Q_k$ in  (\ref{eq:Lyap1}) by
$\Q_k= \F \F^{\dagger}$, to obtain:
\begin{equation}
\hat \psi_k (\omega) = \R( \omega) \F \hat \eta(\omega)~,
\end{equation}
where variables that depend on $\omega$ denote the Fourier amplitudes, i.e.,
\begin{equation}
\hat \psi_k (\omega) = \frac{1}{2 \pi} \int_{-\infty}^{\infty} \hat \psi_k (t) e^{-i \omega t}~ dt~,
\end{equation}
and $\hat \eta(\omega)$ is the Fourier amplitude of the Gaussian noise. The resolvent $\R(\omega)$ 
determines the structure of the response and is given by
\begin{equation}
\R_k (\omega) = (i \omega \I - \A_k)^{-1}~.
\end{equation}
We form the correlation matrix
\begin{equation}
\C_k(\omega) = \left < \hat \psi_k (\omega) \hat \psi_k (\omega)^{\dagger}  \right >  = \R_k(\omega) \Q_k \R_k(\omega)^{\dagger}
\end{equation}
and proceed to calculate the perturbation energy power spectrum as a function of phase velocity as
\begin{equation}
E (c) = \sum_k {\rm trace} \left [ \M_k^{1/2} \C_k ( \omega /k )   \M_k^{1/2} \right ]~,
\end{equation}
and $\M_k$ is the energy metric defined  so that $E_k = \hat \psi_k^{\dagger} \M_k \hat \psi_k$ is the perturbation energy.
The power spectrum is shown in Fig. \ref{fig:10} both for $\kappa=0$ and $\kappa=1$ along with the equivalent normal response
which is obtained by calculating the power spectrum by replacing $\A_k$ by a diagonal matrix with elements its eigenvalues.
If the forcing covariance were the identity the equivalent normal response would be given by the resonance formula:
\begin{equation}
\sum_{k,j} \frac{1}{| i \omega - i \Omega_{k j}|^2}~,
\end{equation}
where $i \Omega_k$ are the eigenvalues of $\A_k$. 
The equivalent normal power spectrum is  equal to the power spectrum when $\A_k$ is a normal  matrix and the forcing is an identity, otherwise
the power spectrum exceeds the equivalent normal power spectrum and the difference reveals the degree of non-normality.
The difference reflects the excess power that is maintained by the system against friction because of the non-orthogonality of the 
eigenmodes \citep{Farrell-Ioannou-1994b, Ioannou-1995}. Note that the power peaks at phase speeds near the maximum and minimum velocity
of the zonal flow. An asymmetry  develops as $\kappa$ increases with power becoming concentrated in the prograde jet
reflecting  the increased instability of the prograde jet as compared with the retrograde jet when
$\kappa>0$.  Because the frequency response arises primarily from the gravest poloidal wavenumber this  double peak in the turbulence spectrum as a function of phase speed is reflected in the frequency spectrum with a double peak at $\omega = k_{min} \overline{v}_{max}$, where
 $k_{min}$ is the poloidal wavenumber corresponding to the gravest mode and $\overline{v}_{max}$ is the maximum velocity of the zonal flow.
 Similar strongly peaked spectra indicative  of coherent large scale  structures  in zonal jet equilibria have been observed \citep{Bush-2003}. 

The particle flux at equilibrium reflects the structures producing it.  
This flux reaches a maximum  as a function of poloidal wavenumber at $m=3$ as seen
in Fig. \ref{fig:7}b. 
The flux  is  downgradient 
where the jet is prograde and becomes upgradient where the jet is retrograde. The difference between the upgradient 
and downgradient particle fluxes leads to a small  downgradient  residual which is
responsible for the eddy energy source.  The regions of upgradient flux show that the particle flux is produced by large 
coherent structures rather than resulting from random advection by small eddies as would be the case  if it were  diffusive.

\subsection{Zonal flow equilibria  for $\kappa >0$}

The dependence  of zonal flow equilibria on the amplitude of the stochastic excitation in the presence of an internal energy   source ($\kappa=1$) 
is similar to that of zonal flows in the case without an internal energy source  ($\kappa=0$).  We find equilibria in the collisionless limit,
$r_m=0$, and these exist for all forcing amplitudes. These equilibria are indicated by a dashed line in the bifurcation
diagram in Fig. \ref{fig:11}c along with the equilibria that result for $r_m=10^{-4}$.  The equilibria for  non zero damping tend to the equilibria
for $r_m =0$ as the stochastic excitation increases.  This asymptotic is reflected in the  eddy induced zonal flow acceleration
which  asymptotes as the stochastic excitation increases (shown in Fig. \ref{fig:11}b).  The eddy kinetic energy at equilibrium increases
with the amplitude of the stochastic excitation and is minimized for zero collisional damping, $r_m=0$.  The particle flux, measured by the average value  $\Gamma_n/ L_x$, increases with stochastic excitation and for zero mean 
collisional damping the particle flux is increasing quadratically with stochastic excitation according to $\Gamma_n / L_x  = 0.0265 u_{rms}^2 $.  From
this it is clear that it would be desirable to operate a device at low stochastic excitation levels and with reduced mean collisional damping if maximizing confinement is the goal. 
All the equilibria of Fig. \ref{fig:11} are structurally  stable for the chosen parameters.  However the basin of attraction of the equilibria is not
the whole space.  Also note that there are no equilibria with
$r_m=10^{-4}$ for   stochastic excitations  smaller than   $u_{rms}=0.065$.

Stochastic excitation, which augments the turbulence,  is important for the equilibration process. In the absence of stochastic excitation the eddy field is dominated 
by the fastest growing modes and the structure of the covariance is not of high enough  rank to comprise the diversity of structures required  to produce
equilibration. At zero or very low stochastic excitation
a vacillation regime is found as occurs for slightly supercritical states in baroclinic turbulence \citep{Pedlosky-77} while for  sufficiently high excitation and associated turbulence levels  one obtains
equilibria. These equilibria for substantial stochastic excitation (i.e. $u_{rms} > 0.1$)
are not only structurally stable but also have a basin of attraction that spans the whole space. However, as the excitation and the supported turbulence is reduced 
the basin of attraction of the equilibria shrinks and finally at a critical value  equilibria cease to exist. 
Operationally, states with low  stochastic excitation and small particle fluxes can be approached by first obtaining an equilibrium by
increasing the stochastic excitation and then adiabatically adjusting the parameters to reach these isolated in parameter space states.

The  vacillation regime mentioned above is not a vacillation of the trajectory of a single realization of the turbulence 
but rather a vacillation regime in the trajectory of the probability density function of the turbulence in the (Gaussian) SSST  approximation.  
%The trajectory results from the influence of turbulent eddy statistics at second order on the jet but it should be born in mind that the SSST dynamics does not account for the influence of the second order statistics of the jet on the turbulence as the ensemble mean of the jet alone is retained in the solution.

We show in  Fig. \ref{fig:12} a state of chaotic DW-ZF fluctuations  at $\kappa=1$  with   very weak forcing [producing  equivalent   $u_{rms} =O(10^{-7}) ~ / (\rho_s \omega_{ci})$ ]  
and zero mean collisional damping. We have determined that there exists an equilibrium state  but this equilibrium state has a small basin of attraction and can not be approached from the SSST initial conditions chosen in this example (which are low turbulence levels, and very small zonal flow).  
A similar chaotic state persists
for these parameters when the  collisional damping is raised to   $r_m=10^{-4}$,  but for that damping  there is no SSST equilibrium  underlying this state
(cf Fig. \ref{fig:11}c).
In Fig. \ref{fig:12}a we see the initial development of the zonal flow, followed by an adjustment period, but unlike
the case with strong stochastic excitation shown in Fig. \ref{fig:2}, which adjusted to equilibrium by stabilizing the perturbations,
the instability remains and alternating  periods ensue  of high eddy activity (low zonal flow)  and low eddy activity (high zonal flow).
The fluctuations settle to a chaotic  bursting pattern  in the zonal flow and the  eddy variables
as  shown in Fig. \ref{fig:13}. The eddy variables, the particle flux at a specific location, the eddy kinetic energy and the integrated
particle flux, have a sawtooth structure in which a slow build up of the eddy variance associated with the underlying instability is followed by a rapid collapse of the eddy fields as the zonal flow develops and  converts the eddy energy to mean zonal energy over 
an advective time scale. The mean zonal kinetic energy exhibits a sawtooth behavior in which  the mean develops very rapidly and then slowly adjusts under the
influence of the weak induced mean eddy  accelerations. Such sawtooth structures have been commonly observed and simulated \citep{Wagner-2007, Diamond-2005}.

For the same parameters that we obtained the chaotic regimes shown  in Fig. \ref{fig:13}  there exists an isolated stable equilibrium with a limited basin of attraction. This equilibrium state can be elicited by impulsively introducing any  SSST zonal flow that is stable at these parameters. Immediately upon
introduction of the zonal flow the eddy energy and the particle flux are quenched and the flow  asymptotically relax to
the equilibrium flow as shown in Fig. \ref{fig:14}. If the parameters do not support an equilibrium DW-ZF state then
the zonal flow  eventually breaks down and a chaotic regime ensues. For example if the collisional damping is raised to $r_m = 10^{-4} \omega_{ci}$ there is no equilibrium at this amplitude of stochastic excitation and in  time $O(1/r_m)$ the imposed zonal flow reverts to a chaotic state.

Regime transition can be controlled using  stochastic excitation as a control parameter.
As the excitation increases the chaotic  bursting gives way to a quasi-periodic regime 
and by further increasing  the stochastic excitation  a fixed point  DW-ZF equilibrium jet state is established as shown in Fig. \ref{fig:15}. 
Having obtained an equilibrium jet state
we then reduce the stochastic excitation (shown in Fig. \ref{fig:16})  and find  that 
the jet persists as the stochastic excitation is reduced and both   the eddy kinetic energy and the particle flux  vanish with the excitation.
This equilibrium state  exists at  the same parameter values for which periodic and chaotic behavior are obtained. 
Hysteretic transition between a steady zonal flow state and a  chaotic turbulent state is  common 
in turbulent systems such as sheared boundary layer flows which   exist in laminar and turbulent states at the same parameter values.

Dependence of zonal flow, eddy variances and fluxes at equilibrium 
on mean collisional damping is shown Fig. \ref{fig:17}; the particle flux increases with mean collisional damping, as does
the eddy energy while the zonal flow velocity decreases, as  is also found in turbulence simulations \citep{Itoh-2007b}.

\subsection{Loss of structural stability at large  $\kappa$}

We now investigate  zonal flow equilibria as a function of $\kappa$. These equilibria, as already discussed,  are most easily initialized
at high stochastic excitation amplitude and low  mean collisional damping.  We study the dependence of these 
equilibria on $\kappa$ at high  turbulence levels (with equivalent $u_{rms}=0.34 ~/ ( \rho_s \omega_{ci} )$).
The maximum zonal flow speed is shown in Fig. \ref{fig:18}a as a function of $\kappa$ and the mean particle flux averaged over the whole domain 
is shown in  Fig. \ref{fig:18}b. The particle flux is seen to initially  increase linearly with $\kappa$. The equilibria are globally attracting up to about $\kappa=1.5$, for the parameters of this problem, but the basin of attraction contracts as $\kappa$ is increased until the flow becomes structurally unstable at  $\kappa=2.534$, and no equilibria exist for larger values of $\kappa$.  Although equilibria exist for $\kappa>1.5$ these equilibria can not be reached from the above listed fixed parameters starting from any initial condition, but they can be  reached
by first establishing an equilibrated state at a lower value of $\kappa$ and then increasing  $\kappa$ adiabatically; although operationally these states are most readily established by  first going to higher
stochastic excitation, corresponding to a higher level of turbulence,  then increasing $\kappa$ and finally reducing  the excitation.

The equilibrated flows and the corresponding minimum damping decay rate of the least damped mode
at each poloidal wavenumber, $m$  are shown in Fig. \ref{fig:19}
for the critical $\kappa_c=2.534$  and for the smaller unstable stratifications  $\kappa=2.52$ and $\kappa=2.0425$. 
As the critical value of $\kappa_c$  is approached the poloidal,   $m=5$ wave,  tends towards instability. 
However,   as $\kappa \rightarrow \kappa_c$ the  damping decay rate of the least damped mode 
approaches, for the chosen parameters
 $kc_{i max} \rightarrow -0.12 \omega_{ci}$ while the fluxes and the equilibrium zonal flow tend to diverge as  $\kappa_c$ is approached and 
no equilibrium flows can be sustained for $\kappa > \kappa_c$ and transition to a time varying state occurs.  This result shows that the jet first loses structural stability as a function of $\kappa$ rather than modal stability.

\section{Discussion}

There are a number of points we wish to emphasize in connection with the above results:

\begin{enumerate}
\item A novel concept arising from SSST is that of the structural stability boundary for zonal flow breakdown as distinct from breakdown related to shear instability of the zonal flow. 

\item Multiple DW-ZF regimes are predicted to exist in  parameter space including a regime of steady zonal flows as well as  regimes of periodic, quasi-periodic and  chaotic bursting or ``sawtooth" behavior. These regimes provide  opportunity for placing  and manipulating confinement devices to be in a desired  dynamical state between high and low confinement.

\item  SSST  predicts that isolated DW-ZF equilibria at high $\kappa$   are not connected continuously to lower $\kappa$ states but that these states can be reached either using external turbulence excitation or finite amplitude state perturbation to promote the system between these equilibria.

\item A mechanism for introducing and modulating turbulence levels is predicted to provide a powerful control parameter for placing the DW-ZF system in  desired confinement states.

\item In the limit of vanishing zonal flow collisional damping a universal DW-ZF state is supported in which a precise balance between down gradient momentum transport by small wavenumbers and upgradient transport by high zonal wavenumbers occurs. This asymptotic equilibrium predicts that  band limiting turbulence can prevent formation of stable equilibrium zonal flows.

\item Density fluxes are not diffusive but rather are primarily produced by large scale structures. Robust fluxes both up and down the mean gradient occur and it follows that particle transport analysis requires a full wave solution. 

\end{enumerate}

 \section{Conclusion}

Emergence of zonal jets in turbulent flow and the relation of  these jets to the statistical equilibrium of the turbulent state is a problem of great theoretical and practical interest.  This problem is particularly compelling  in the case of turbulent plasmas because of the relationship of zonal jets to the  H states that limit  turbulent transport of particles and heat in magnetic confinement fusion devices.  DW-ZF interaction dynamics is responsible for the generation and regulation of these zonal flows so it follows that prospects for  predicting and controlling the H state require  improvement in  fundamental understanding of the mechanism underlying the statistical steady state of zonal jets in drift wave turbulence.   In this work we applied the methods of SSST  to the  
Hasegawa-Wakatani model  to study the emergence, stability  and effect on transport of zonal jets in the DW-ZF system.  We find robust zonal jet formation in agreement with both experiment and simulation and obtain parameter requirements for jet formation and breakdown. We find multiple regimes including
chaotic, periodic and steady and show that externally imposed turbulence and finite amplitude zonal flow perturbations can be used to control regime transition.   We find suppression of particle transport by zonal flows and show that this transport is not diffusive in nature.  These results provide a basis for prediction and controlling confinement regimes in DW-ZF turbulence.

\begin{acknowledgment}
Discussions with S.Nazarenko and B. Nadiga are gratefully acknowledged. 
This work was partially supported  by NSF ATM-0123389. 
\end{acknowledgment}
\clearpage

\newpage 
\bibliographystyle{ametsoc}
\bibliography{basic_references_20090803}
%\begin{figure}[h]
%\centering
%\includegraphics[width=8in]{transition.eps}
% \vspace*{-1mm} \caption{ }
% \label{fig:1}
%\end{figure}
\clearpage

\begin{figure}[h]
\centering
\includegraphics[width=6in]{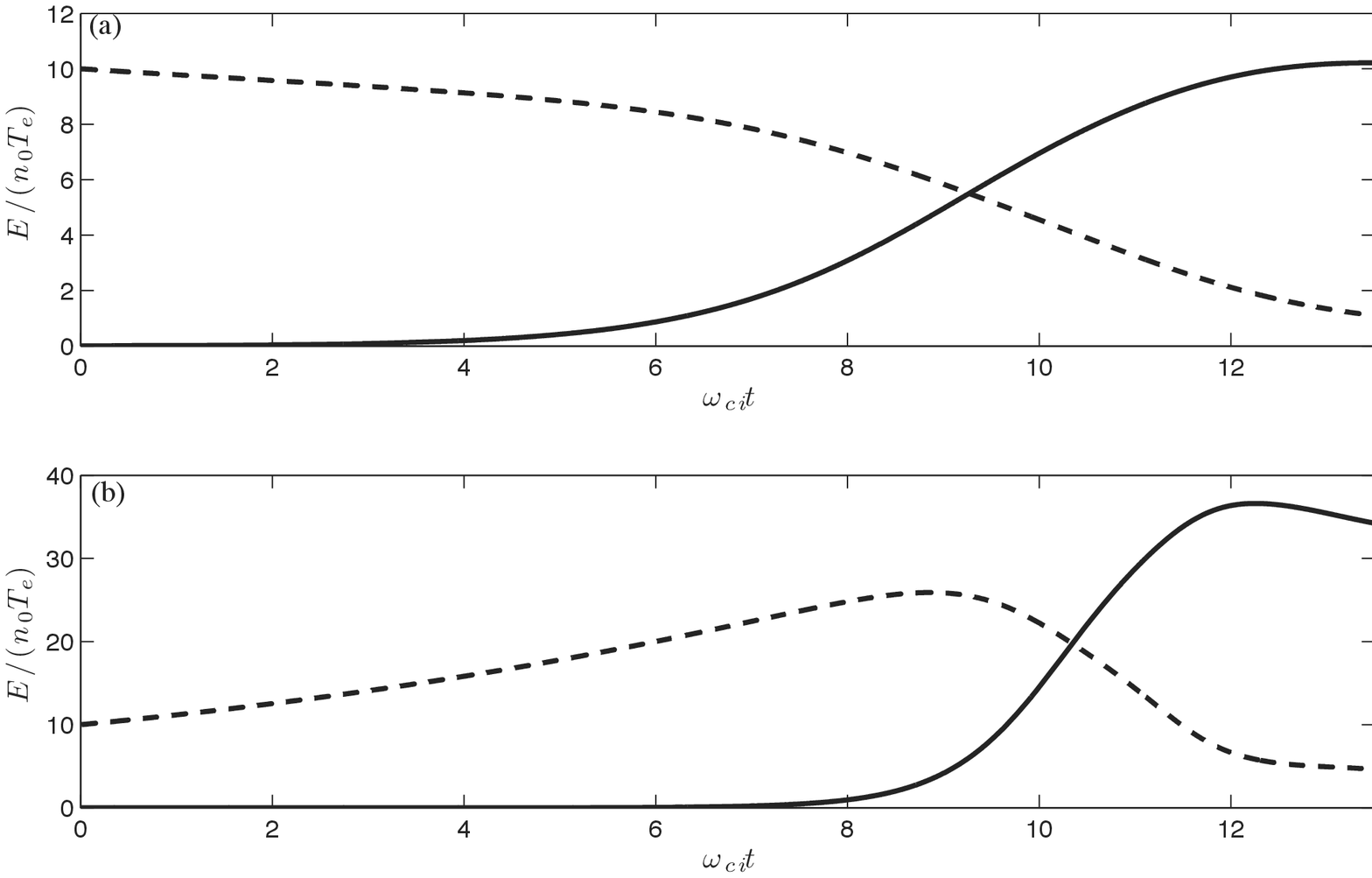}
 \vspace*{-1mm} \caption{ Initial jet formation by the rapid adjustment process starting from a state of strong turbulence for the cases  (a) $\kappa =0$ (no instability) and  (b) $\kappa=1$ (strong instability).   Shown are eddy kinetic energy (dashed) and mean zonal kinetic energy (solid) as a function of time.  The eddy field is limited to global zonal wavenumber  $m=5$ and there is no stochastic excitation.}
 \label{fig:1}
\end{figure}

\clearpage

\begin{figure}[h]
\centering
\includegraphics[width=6in]{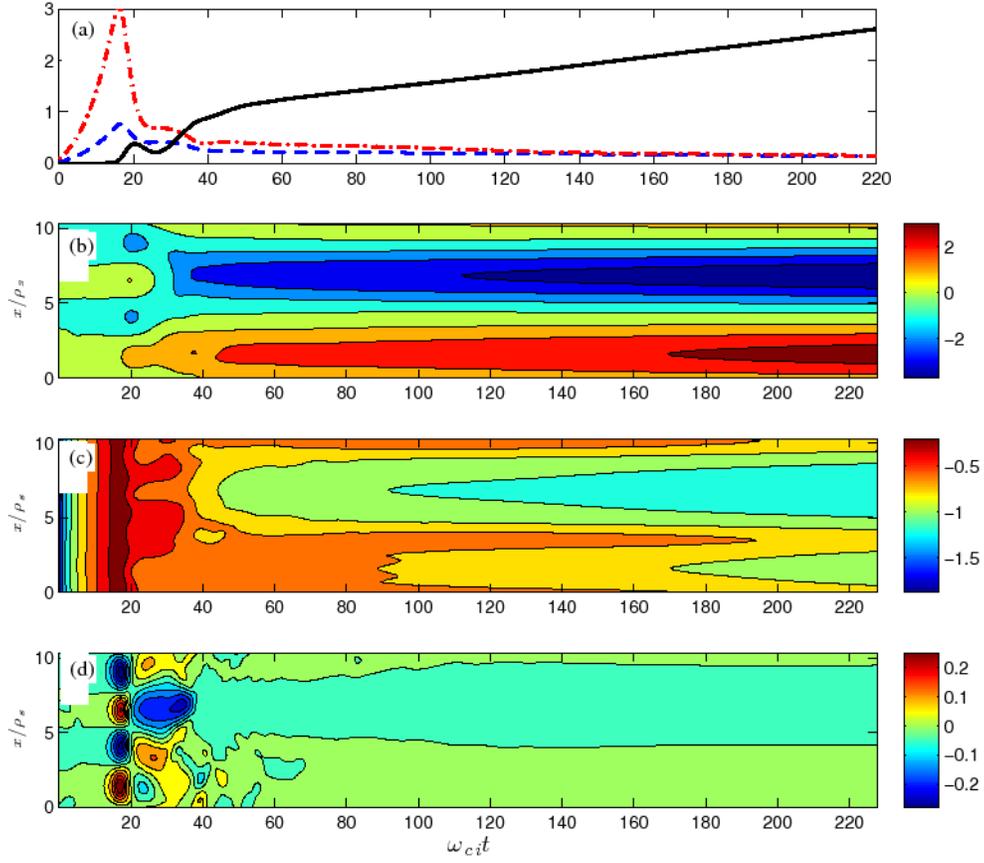}
 \vspace*{-1mm} \caption{Transient development of an equilibrium zonal jet. (a) time development of
 the  mean kinetic energy of the zonal flow, $E_m / (n_0 T_e)$ (solid), the mean eddy kinetic energy $K_e / (n_0 T_e)$  (dashed) and the
 total particle flux over the channel $\Gamma_n \omega_{ci} / (n_0 T_e)$ (dash-dot). 
 (b): zonal velocity, $V / (\rho_s \omega_{ci})$, as a function of the radial direction  and time. 
 (c): eddy kinetic energy,  $\log_{10}(K_e/(n_0T_e))$,  as a function of the radial direction  and time.
 (d): eddy induced zonal flow  acceleration, $-< u' \zeta' > / (\rho_s \omega_{ci}^2)$,  
 as a function of the radial direction  and time.  The parameters are: $\kappa=1$,  $r_m=10^{-4} \omega_{ci}$
 and the stochastic excitation  has  equivalent r.m.s. velocity of $0.34 \rho_s \omega_{ci}$.} 
\label{fig:2}
\end{figure}

\clearpage
\begin{figure}[h]
\centering
\includegraphics[width=6in]{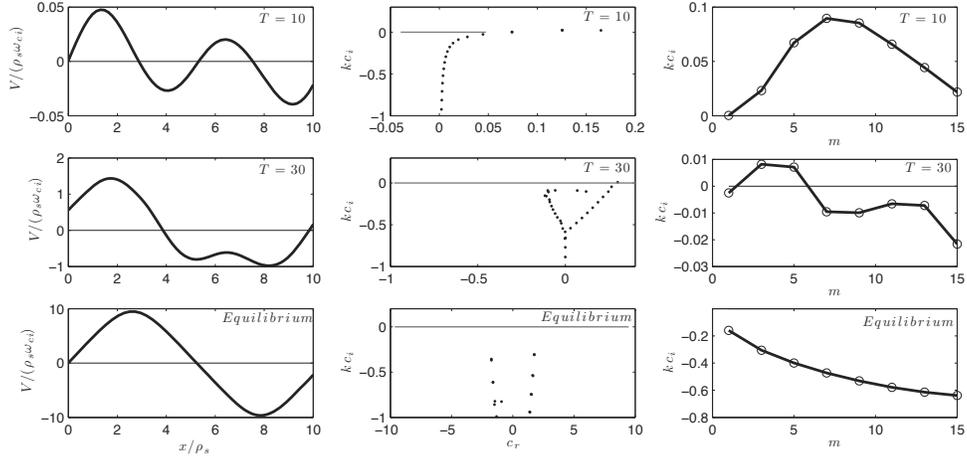}
 \vspace*{-1mm} \caption{Evolution of the zonal flow and its associated spectrum for the example in Fig. 2.  Left panels: zonal flow structure
 at $T=10, 30$ and at equilibrium. Center panels: spectrum $(c_r, k c_i)$  of   $\A_k$ for
 the flow  in the corresponding panel for zonal wavenumber $m=3$.  
 The continuous line indicates  the velocity interval spanned by the zonal flow.  At equilibrium the instabilities  have been stabilized.
 Right panels: the largest growth rate for a given zonal wavenumber, $m$. 
 At equilibrium the  least stable mode corresponds to the gravest zonal wavenumber.}
 \label{fig:3}
\end{figure}

\clearpage
\begin{figure}[h]
\centering
\includegraphics[width=6in]{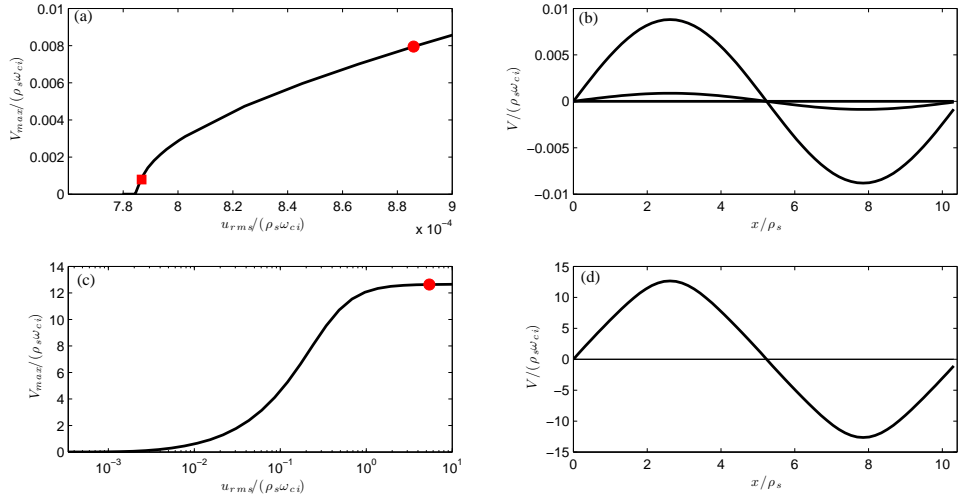}
 \vspace*{-1mm} \caption{(a) Maximum  zonal flow   velocity as a function of  stochastic excitation for $\kappa=0$. Stochastic excitation
 is measured by the  $u_{rms}$  that would have been maintained  in the absence  flow. For the chosen  parameters the critical forcing required to form zonal flows
 is  $u_{rms} = 7.8 \times 10^{-4} / (\rho_s \omega_{ci} )$.
 (b) Corresponding equilibrium  zonal zonal flows: the larger velocity corresponds to forcing denoted with a circle in panel (a), while the smaller velocity
 corresponds to the parameters denoted with a square in (a).
 (c) Continuation of the bifurcation diagram of (a) to larger forcing values. Note that as the forcing increases the maximum zonal flow velocity asymptotes to a constant.
 (d) The asymptotic zonal flow  at large forcing. 
 The collisional damping of the mean is $r_m = 10^{-4} \omega_{ci}$.}
 \label{fig:4}
\end{figure}

\clearpage
\begin{figure}[h]
\centering
\includegraphics[width=6in]{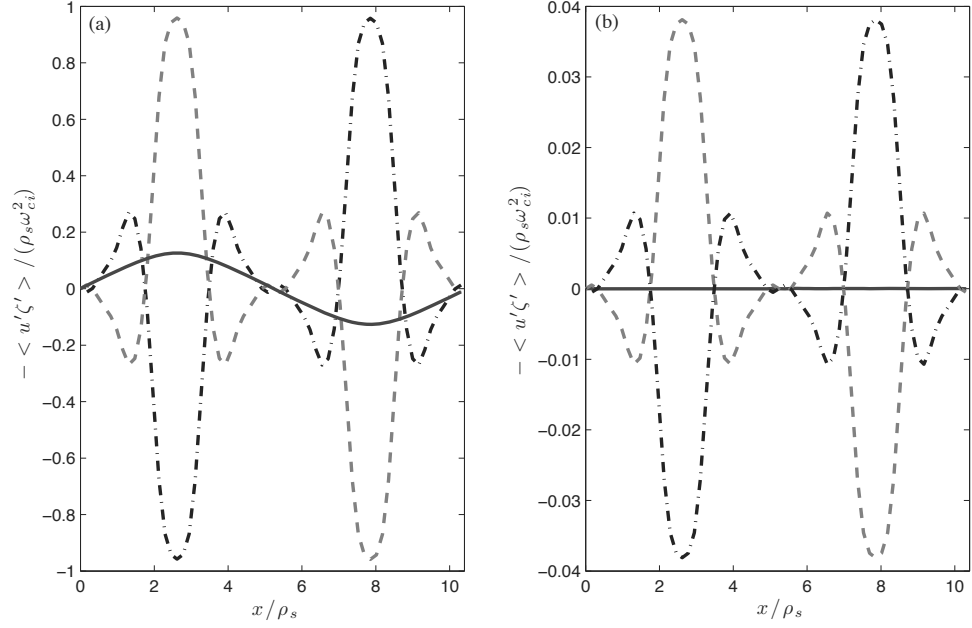}
 \vspace*{-1mm} \caption{Structure of the  eddy induced zonal flow  acceleration $-<\overline{u' \zeta'} > $
 as a function of radius.
 The solid line is the total flux summed over all zonal  wavenumbers multiplied by 100 (at equilibrium this is equal to $100 r_m \overline{v}$).
 The dashed line is the acceleration induced by wavenumbers $m=7-15$. These higher waves  build the zonal flow.
 The dash-dot line is the acceleration induced  by the small wavenumbers  $m=1-5$ which tend to destroy the zonal flow.
 Left panel: for mean collisional damping $r_m=10^{-4}$ and the  equilibrium flow in Fig. \ref{fig:1}d. 
 Right panel: For $r_m=0$ (here the cancellation between 
 downgradient and upgradient
 fluxes is perfect).}
 \label{fig:5}
\end{figure}

\clearpage

\begin{figure}[h]
\centering
\includegraphics[width=6in]{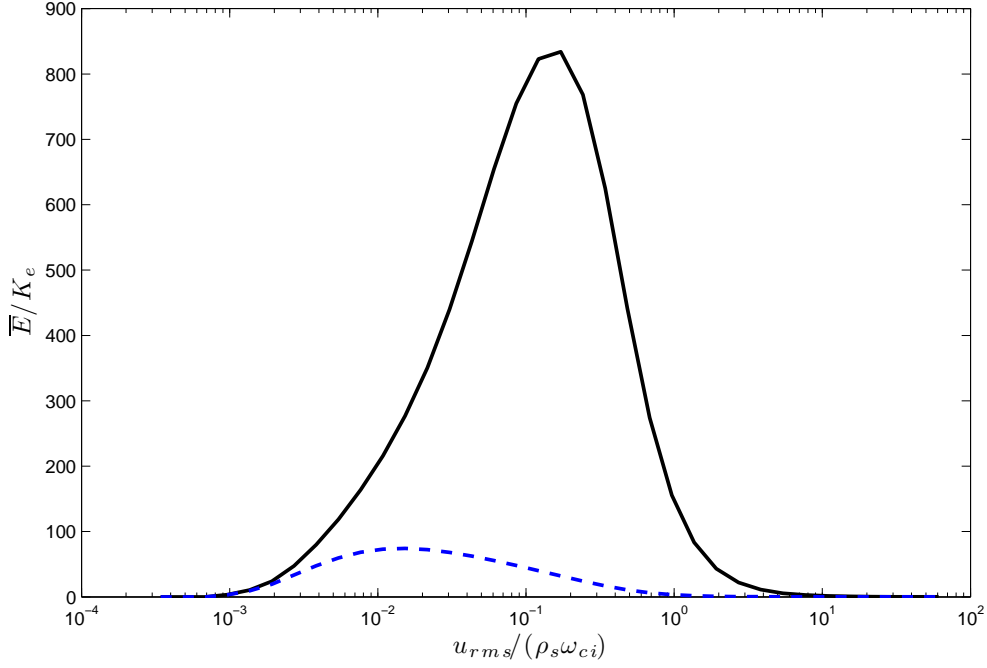}
 \vspace*{-1mm} \caption{Ratio of mean zonal kinetic energy to eddy kinetic energy as a function of stochastic excitation (solid).
 Ratio of the mean zonal kinetic energy to the eddy kinetic that would have been maintained in the absence of  the zonal flow (dashed).
 For small excitations there is no zonal flow and the ratio vanishes, also for large excitations the flow asymptotes to a constant and again the ratio
 vanishes. For intermediate excitations the zonal flow energy is  two to three orders of magnitude larger and the turbulence energy  is dominated by the 
 zonal flow energy.  The zonal flow suppresses the eddy energy by approximately an order of magnitude. For $\kappa=0$ and $r_m= 10^{-4} \omega_{ci}$.} 
 \label{fig:6}
\end{figure}

\clearpage
\begin{figure}[h]
\centering
\includegraphics[width=6in]{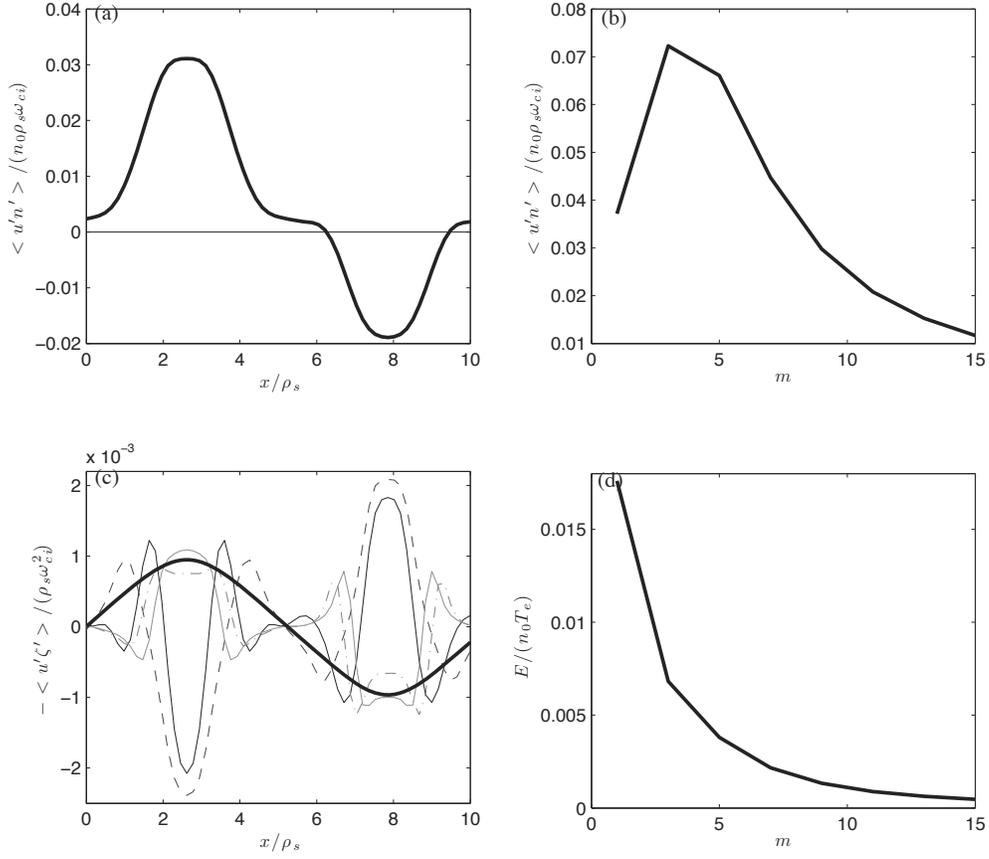}
 \vspace*{-1mm} \caption{(a) Structure in radius of the particle flux at equilibrium. The particle flux is not diffusive, as it has a distinct structure and 
there is a region of upgradient flux that would correspond to a negative diffusion coefficient.  (b)  The integrated particle flux at equilibrium as a function
of zonal wavenumber, $m$. (c) The structure of the eddy acceleration  $-<u' \zeta' >$  produced by the  zonal modes. The thick solid line is the total 
vorticity flux  which maintains  the zonal flow  against dissipation shown in Fig. \ref{fig:8}.  The opposing fluxes (solid and dashed) is the flux associated with
wavenumbers $m=1,3$ while the supporting fluxes (solid and dashed-dot) correspond to the higher wavenumbers $m=5,7$. 
(d) The energy of the eddy field as a function of zonal wavenumber.  The eddy kinetic energy peaks at the gravest zonal mode  $m=1$.
The case is for $\kappa=1$ $r_m = 10^{-4} ~\omega_{ci}$ and stochastic excitation equivalent to
r.m.s. velocity of $0.34  \rho_s \omega_{ci}$.}
 \label{fig:7}
\end{figure}

\clearpage
\begin{figure}[h]
\centering
\includegraphics[width=6in]{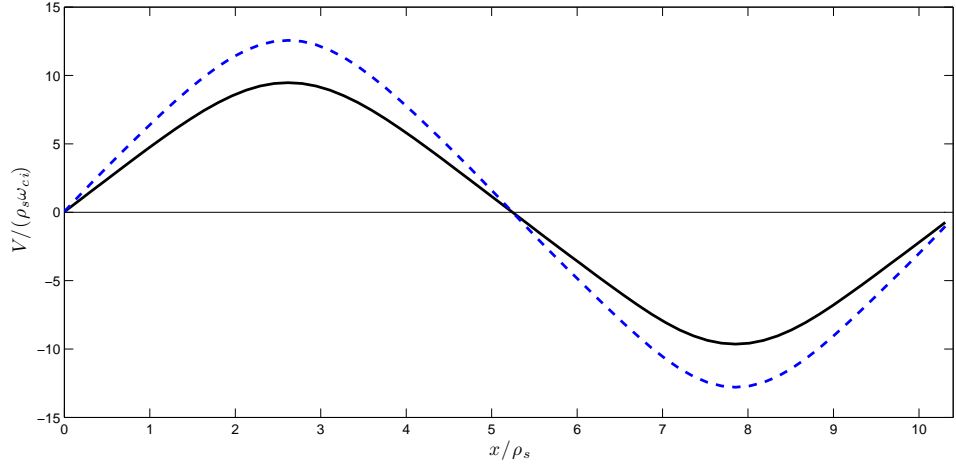}
 \vspace*{-1mm} \caption{Zonal flow at equilibrium as a function of radius.
 Dashed: with no collisional damping of the mean ($r_m=0$);  solid: with $r_m = 10^{-4} \omega_{ci}$.
 The case is for $\kappa = 1$, and stochastic forcing with equivalent r.m.s. velocity of $0.34 \rho_s \omega_{ci}$.}
 \label{fig:8}
\end{figure}

\clearpage
\begin{figure}[h]
\centering
\includegraphics[width=6in]{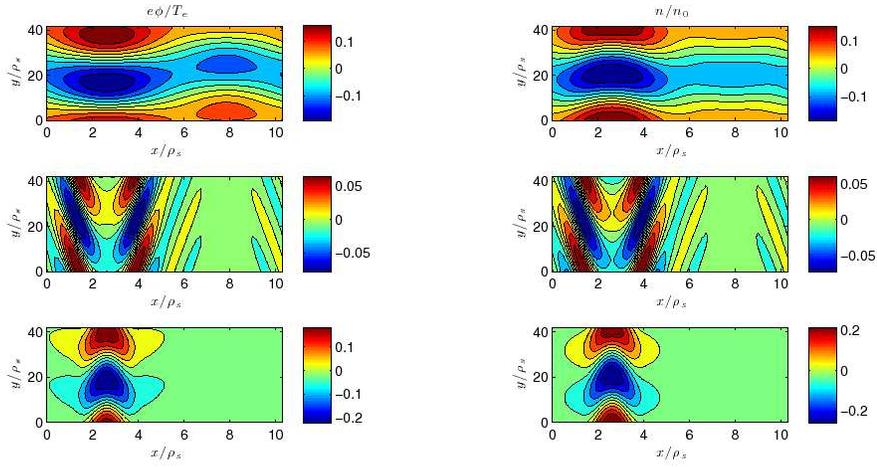}
 \vspace*{-1mm} \caption{(Color online) Top row: the top EOF of the eddy covariance of the component of the  eddy field with 
 zonal wavenumber $m=1$  (on the left: perturbation 
 electric field, on the right:   perturbation density). The first EOF accounts for $32\%$ of the total energy of the eddy field
 at this wavenumber. Middle row: the stochastic optimal. The stochastic optimal produces
 $20 \%$ of the eddy energy at this wavenumber. Bottom row: the least stable eigenvalue of the operator
 at $m=1$. The associated growth rate is $k c_i = -0.15 \omega_{ci}$. For the equilibrium  zonal flow obtained at $\kappa=1$ 
 with stochastic excitation equivalent to
equivalent r.m.s. velocity of $0.34 \rho_s \omega_{ci}$ }
 \label{fig:9}
\end{figure}

\clearpage
\begin{figure}[h]
\centering
\includegraphics[width=6in]{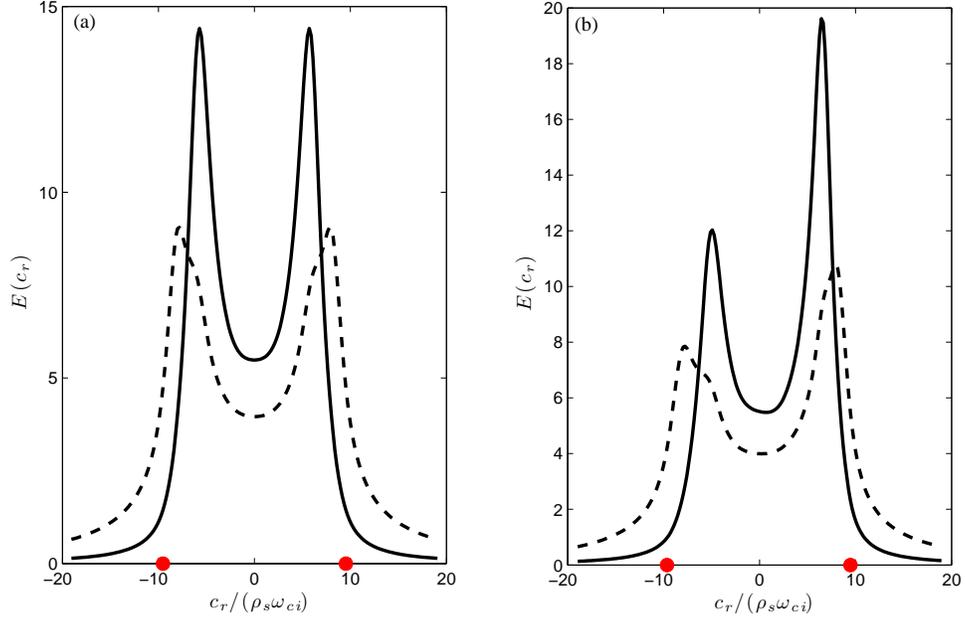}
 \vspace*{-1mm} \caption{(Color online) Power spectrum of the eddy energy as a function of phase speed $c_r$ (solid).
The dashed line is the equivalent normal response and circles mark the maximum and minimum velocity of the equilibrium flow. (a) for
 $\kappa=0$.   (b) for $\kappa=1$.  
 The case is for equivalent r.m.s. velocity of $0.34 \rho_s \omega_{ci}$ and $r_m=10^{-4}$. }
 \label{fig:10}
\end{figure}

\clearpage
\begin{figure}[h]
\centering
\includegraphics[width=6in]{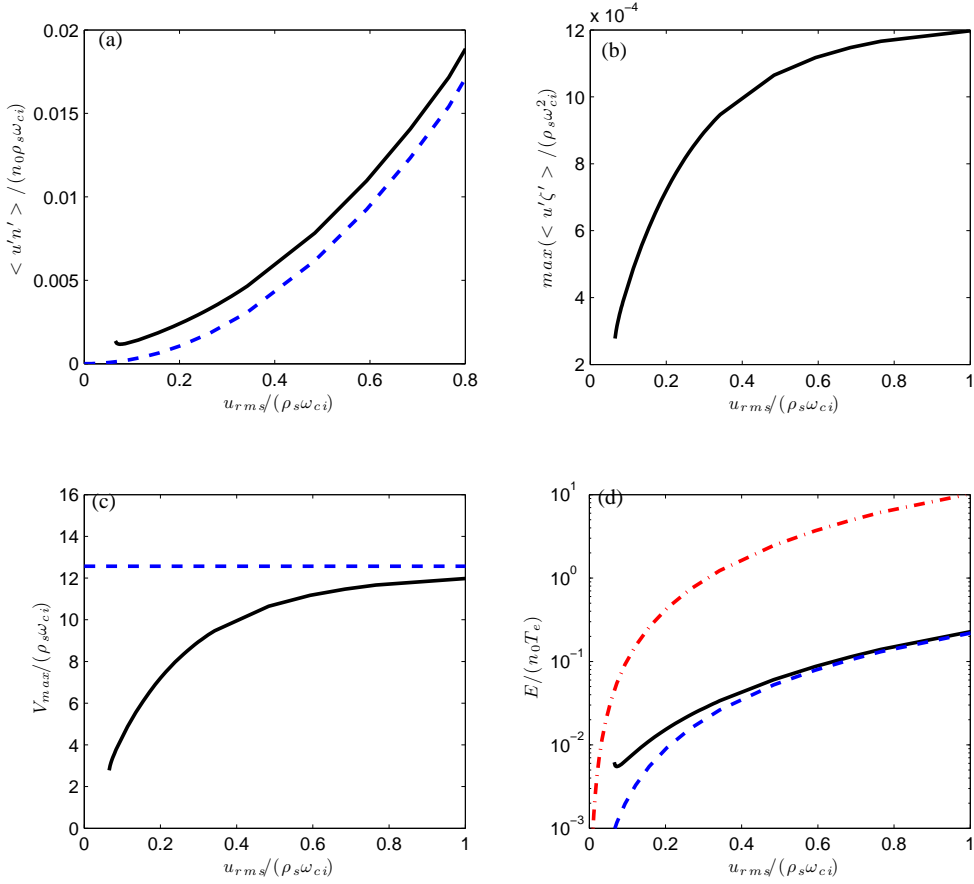}
 \vspace*{-1mm} \caption{(Color online) (a) Particle flux as a function of  stochastic excitation measured by equivalent $u_{rms}$;  for $r_m= 10^{-4} \omega_{ci}$ (solid) and for
 $r_m=0$ (dashed).
(b) Maximum vorticity flux $ < u' \zeta' > $  as a function of  stochastic excitation. 
(c) Maximum equilibrium zonal flow velocity as a function of  stochastic excitation; for $r_m= 10^{-4} \omega_{ci}$ (solid) and for
 $r_m=0$ (dashed).
(d) Mean eddy kinetic energy as a function of stochastic excitation.   Also shown is the   eddy kinetic energy 
maintained against dissipation  in the absence of flow as a function of stochastic excitation (dashed-dot).
 For $\kappa = 1$.}
 \label{fig:11}
\end{figure}

\clearpage

\begin{figure}[h]
\includegraphics[width=6in]{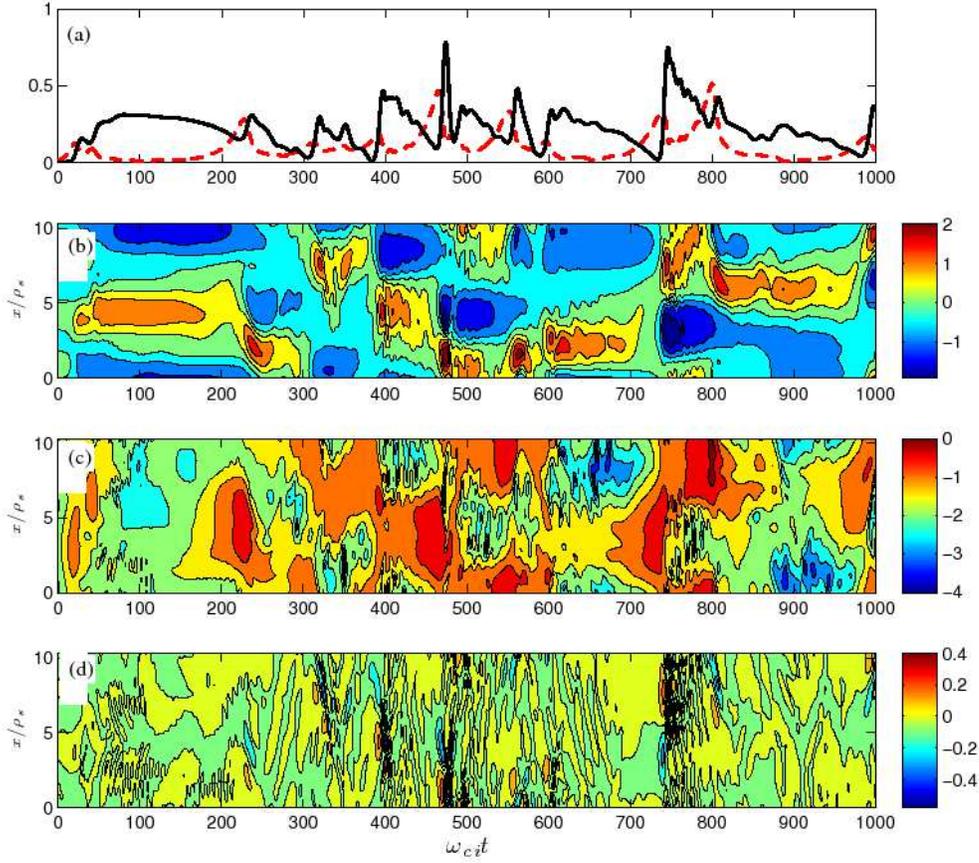}
 \vspace*{-1mm} \caption{A  chaotic state (analysis of perturbed trajectory differences reveals this system to be chaotic with Lyapunov exponent $0.02 \omega_{ci}$). (a):  Zonal flow energy (solid), and eddy kinetic energy (dashed)  as a function of time.
 (b): zonal velocity, $V / (\rho_s \omega_{ci})$,  as a function of  radius and time. 
 (c): eddy kinetic energy, $\log_{10}(K_e/(n_0T_e))$, as a function of  radius  and time.
 (d): eddy induced zonal flow  acceleration, $-< u' \zeta' > / (\rho_s \omega_{ci}^2)$,  
 as a function of  radius and time.  The parameters are: $\kappa=1$,  $r_m=0$
 and the stochastic excitation  has  equivalent r.m.s. velocity of $0.34 \times 10^{-7} \rho_s \omega_{ci}$.
 For these values there exists an equilibrium zonal flow with a limited  basin of attraction, 
 and this equilibrium state can not be approached from initial states with small zonal flows.}
 \label{fig:12}
\end{figure}

\clearpage
\begin{figure}[h]
\centering
\includegraphics[width=6in]{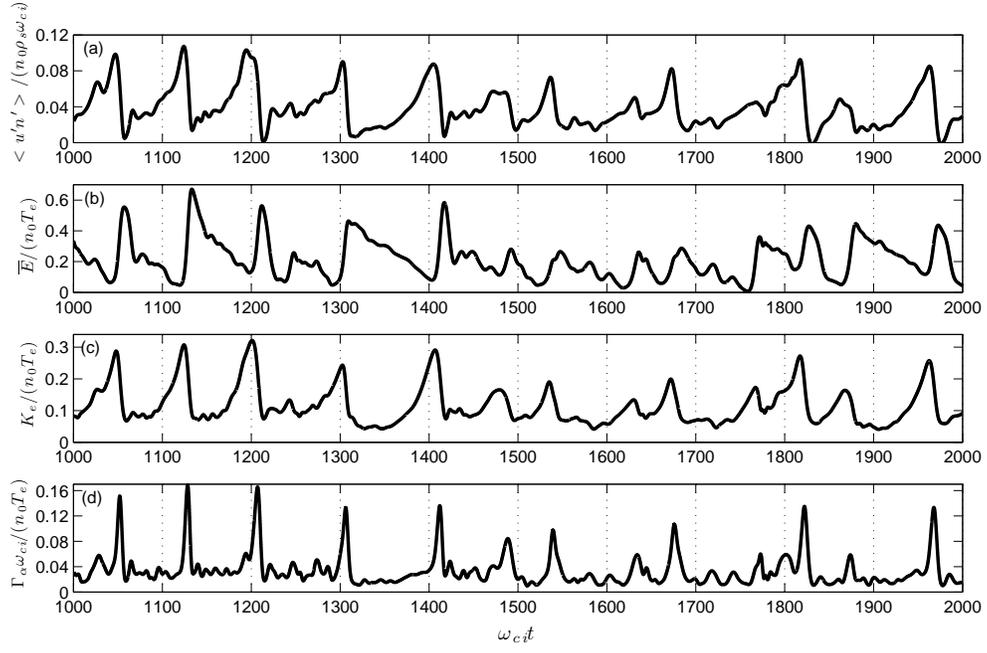}
 \vspace*{-1mm} \caption{For the case shown in Fig. \ref{fig:12}: 
 (a) particle flux at a single location as a function of time; 
 (b) zonal flow kinetic energy;
 (c) eddy kinetic energy;
 (d) average  particle flux. 
 } 
 \label{fig:13}
\end{figure}

\clearpage
\begin{figure}[h]
\begin{center}
\includegraphics[width=6in]{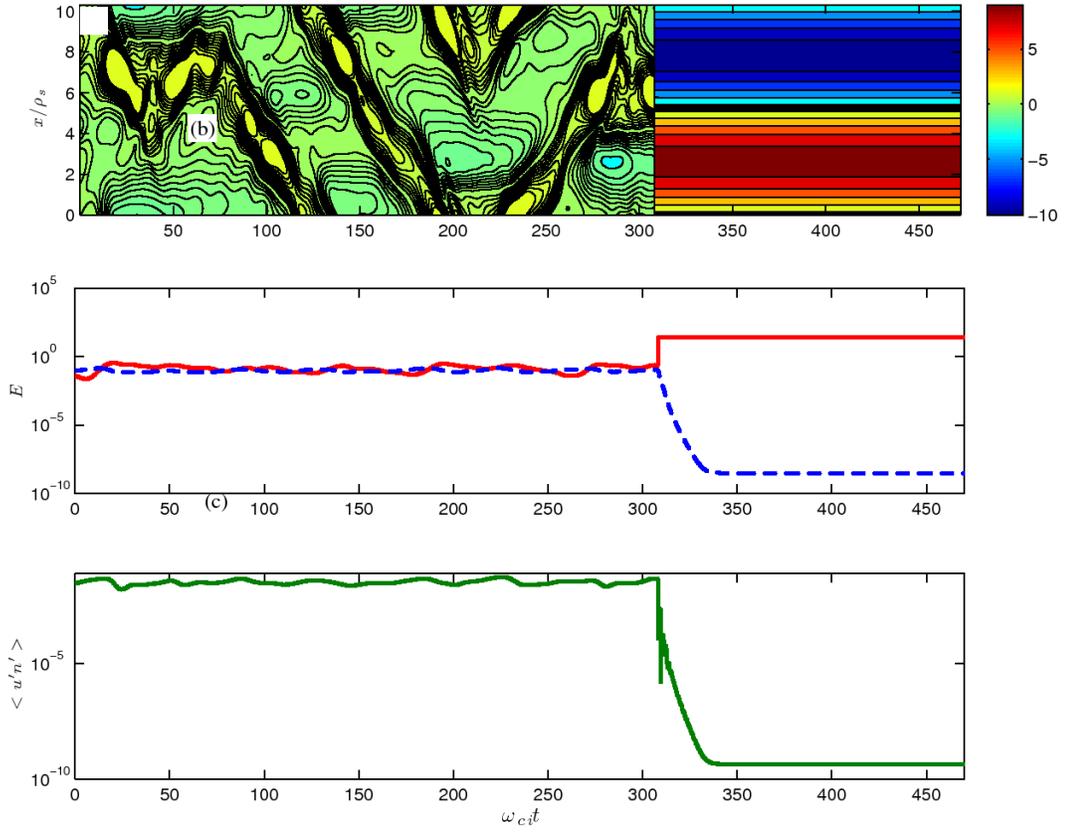}
 \vspace*{-1mm} \caption{A  chaotic state is laminarized by  impulsive introduction of a stable zonal flow at $\omega_{ci} t= 310$. The zonal flow subsequently asymptotically approach the equilibrium zonal flow that exists for these parameter values.
  (a): zonal velocity, $V / (\rho_s \omega_{ci})$,  as a function of  radius and time. 
   (b):  Zonal flow energy (solid), and eddy kinetic energy (dashed)  as a function of time.
    (c):  Mean particle flux as a function of time
For $\kappa=1$ and $r_d= 0$.}
 \label{fig:14}
 \end{center}
\end{figure}

\clearpage

\begin{figure}[h]
\begin{center}
\includegraphics[width=6in]{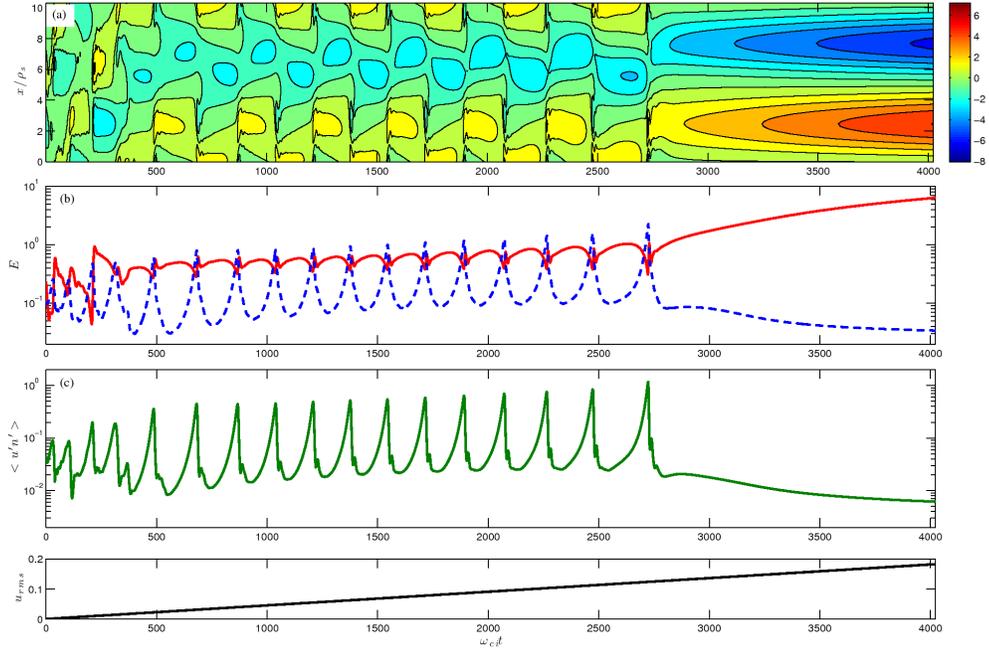}
 \vspace*{-1mm} \caption{A  chaotic state $(0<\omega_{ci} t < 400$)  becomes quasi periodic $(450<\omega_{ci} t < 2800$)  and then settles to an equilibrium as  stochastic excitation increases ($0.34 \times 10^{-7} \rho_s \omega_{ci}< u_{rms} <0.1823 \rho_s \omega_{ci}$).
  (a): zonal velocity, $V / (\rho_s \omega_{ci})$,  as a function of  radius and time. 
   (b):  Zonal flow energy (solid), and eddy kinetic energy (dashed)  as a function of time.
    (c):  Mean particle flux as a function of time
    (d)  Stochastic excitation as a function of time.
For $\kappa=1$ and $r_d=0 $.}
 \label{fig:15}
 \end{center}
\end{figure}
\clearpage

\begin{figure}[h]
\begin{center}
\includegraphics[width=6in]{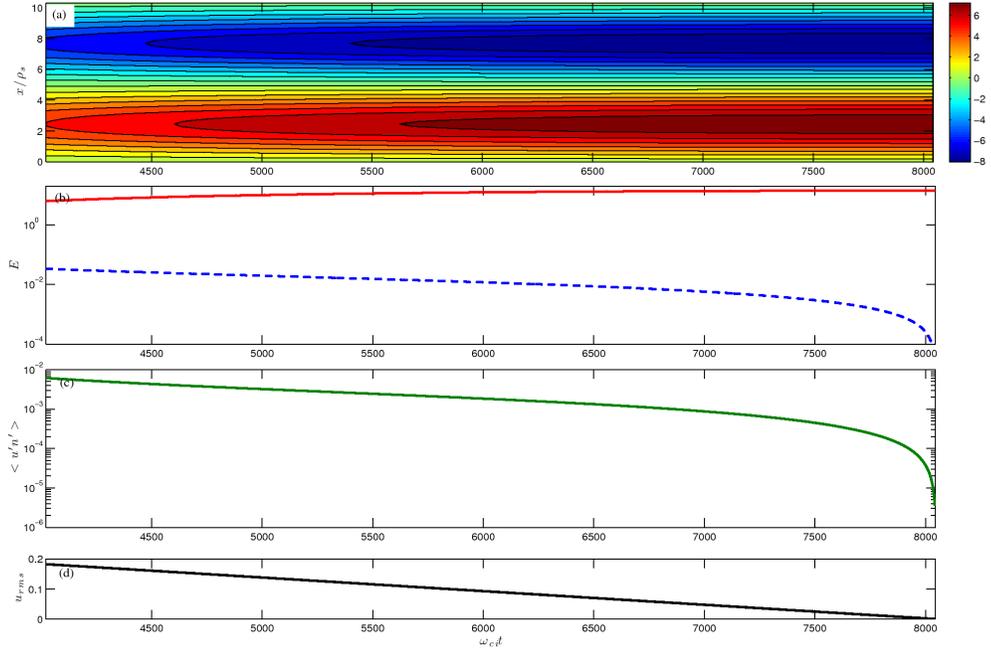}
 \vspace*{-1mm} \caption{Continuation of Fig. \ref{fig:15}. The stochastic excitation is decreased to its initial value 
 ($u_{rms}=0.34 \times 10^{-7} \rho_s \omega_{ci}$). The zonal flow persists while  the eddy
 kinetic energy and the particle flux  vanish with the excitation.
  (a): zonal velocity, $V / (\rho_s \omega_{ci})$,  as a function of  radius and time. 
   (b):  Zonal flow energy (solid), and eddy kinetic energy (dashed)  as a function of time.
    (c):  Mean particle flux as a function of time
    (d)  Stochastic excitation as a function of time.
For $\kappa=1$ and $r_d=0$.}
 \label{fig:16}
 \end{center}
\end{figure}

\clearpage

\begin{figure}[h]
\centering
\includegraphics[width=6in]{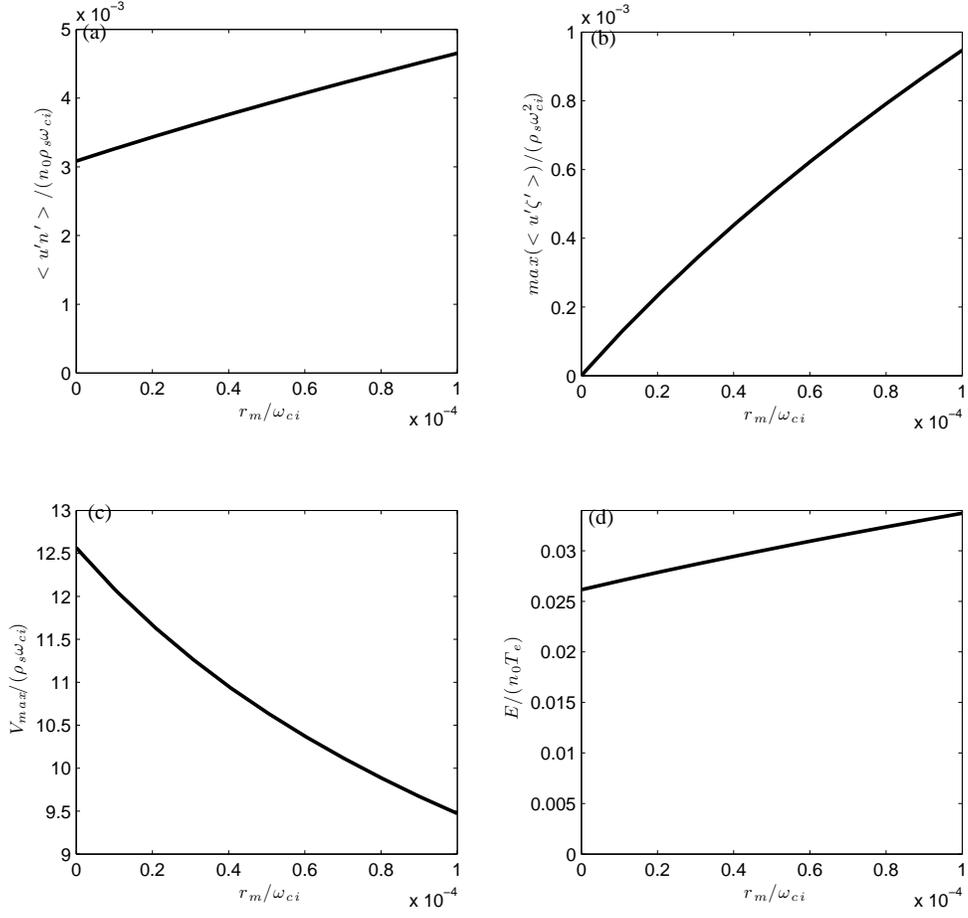}
 \vspace*{-1mm} \caption{Equilibrium state diagnostics as a function of mean collisional damping. (a) Particle flux. (b) 
 Maximum vorticity flux. (c)
 Maximum equilibrium zonal flow velocity. 
 (d) Mean eddy kinetic energy.
The case is for $\kappa = 1$, and stochastic forcing with equivalent r.m.s. velocity of $0.34 \rho_s \omega_{ci}$.}
 \label{fig:17}
\end{figure} 
\clearpage

\begin{figure}[h]
\centering
\includegraphics[width=6in]{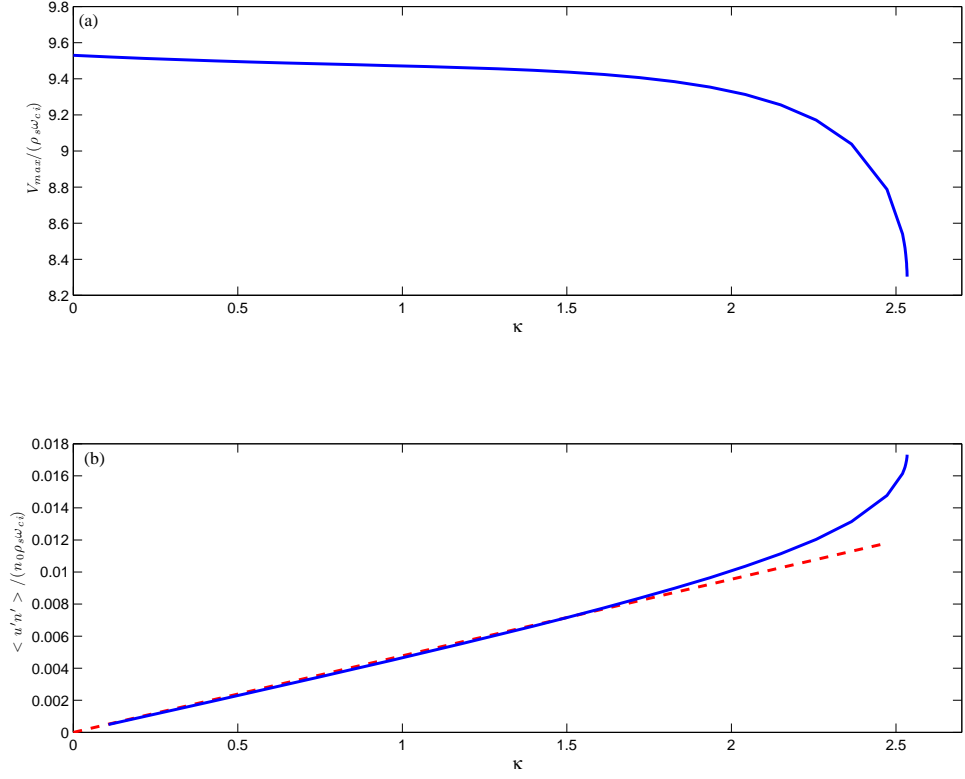}
 \vspace*{-1mm} \caption{Equilibrium state diagnostics as a function of density gradient, $\kappa$ (a): Maximum velocity of the equilibrium zonal flow.  
% The corresponding r.ms. eddy
% velocity  is between $25$ times smaller at $\kappa=0$ and $20$ fold smaller at $\kappa=2.5$  just before the critical  structural instability point.
 (b) The mean particle flux  (solid). The mean particle flux increases at first linearly as $0.05 \kappa /Lx $ (dashed).   
 The parameters are:  $r_m=10^{-4}$
 and the stochastic excitation supports  equivalent r.m.s. velocity of $0.34  \rho_s \omega_{ci}$.}
 \label{fig:18}
\end{figure}
\clearpage

\begin{figure}[h]
\centering
\includegraphics[width=6in]{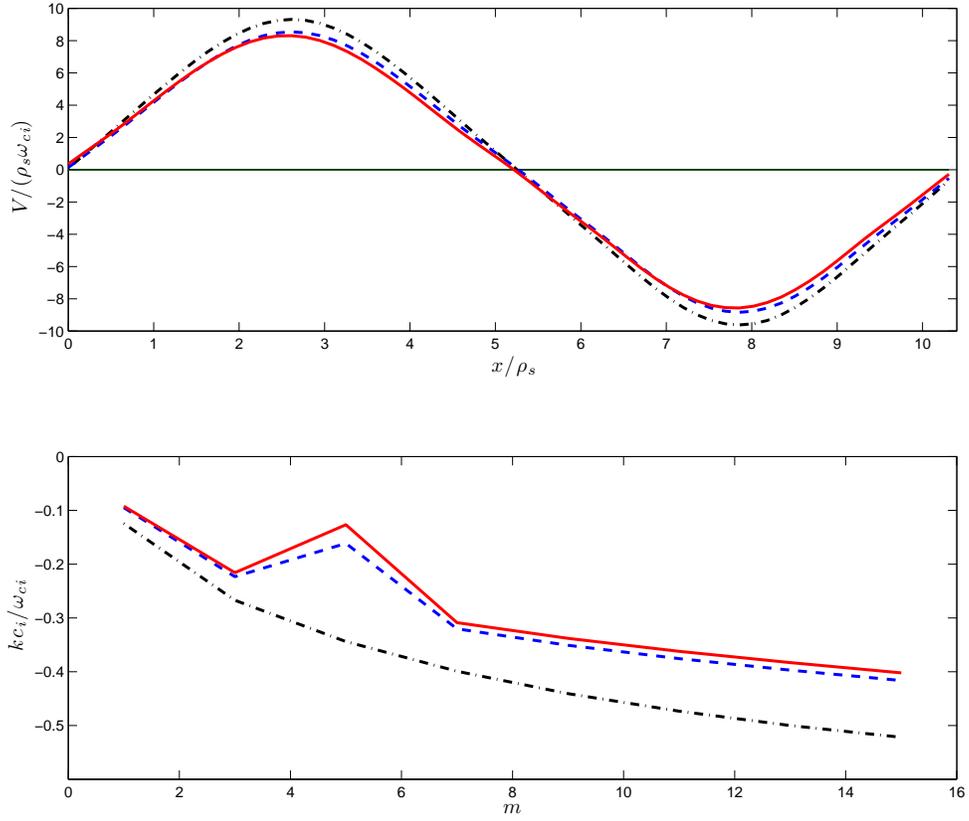}
 \vspace*{-1mm} \caption{Approach to structural instability as a function of $\kappa$. Top: Zonal flow velocities as the critical $\kappa_c=2.534$ is approached. 
 Bottom: The corresponding  maximum growth rate of perturbations  as a function of poloidal wavenumber, $m$. Solid: $\kappa=2.534$, dash: $\kappa=2.52$, dash-dot:   for $\kappa=2.0425$. The parameters are:  $r_m=10^{-4}$ and the stochastic excitation  has  equivalent r.m.s. velocity of $0.34  \rho_s \omega_{ci}$.} 
 \label{fig:19}
\end{figure}

\end{document}